\documentclass[aps,prx,twocolumn,showpacs,10pt]{revtex4-1}
\usepackage{graphicx}
\usepackage{amssymb}
\usepackage{amsmath,color,mathrsfs}
\usepackage{bbold}
\usepackage{comment}
\usepackage[dvipsnames]{xcolor}
\usepackage{bm}
\pdfoutput=1
\usepackage{hyperref}
\hypersetup{urlcolor=blue, colorlinks=true, citecolor=blue, linkcolor=blue}

\usepackage{todonotes}

\newcommand{\up}{\uparrow}
\newcommand{\down}{\downarrow}

\begin{document}

\title{Finite-temperature conductance of strongly interacting quantum wire with a nuclear spin order}

\author{Pavel P. Aseev}
\author{Jelena Klinovaja}
\author{Daniel Loss}
\affiliation{Department of Physics, University of Basel, Klingelbergstrasse 82, CH-4056 Basel, Switzerland}


\begin{abstract}

We study the temperature dependence of the electrical conductance of a clean strongly interacting quantum wire in the presence of a helical nuclear spin order. The nuclear spin helix opens a temperature-dependent partial gap in the electron spectrum. Using a bosonization framework we describe the gapped electron modes by sine-Gordon-like kinks. We predict an internal resistivity caused by an Ohmic-like friction these kinks experience via interacting with gapless excitations. As a result, the conductance rises from $G=e^2/h$ at temperatures below the critical temperature when nuclear spins are fully polarized to $G=2e^2/h$ at higher temperatures when the order is destroyed, featuring a relatively wide plateau in the intermediate regime. The theoretical results are compared with the experimental data for GaAs quantum wires obtained recently by Scheller~\textit{et~al.}~[Phys.~Rev.~Lett.~\textbf{112},~066801~(2014)].

\end{abstract}

\maketitle

\section{Introduction}

Basic electronic properties of three-dimensional (3D) interacting electron systems are usually well described
within the Landau Fermi-liquid picture where low-energy excitations are single-electron
quasiparticles. This is not the case in 1D systems where interaction cannot be considered as a small perturbation. Rather than electronic quasiparticles, the low energy excitations are collective density waves~(bosons), and the system can be described as a Luttinger liquid~(LL)~\cite{GiamarchiBook, VoitRepProg1995}.

In  recent years, helical and quasi-helical LLs, special classes of LLs exhibiting spin-filtered transport, have received much attention. The helical LL describes, for example, edges of two-dimensional topological insulators \cite{KonigJPhysSocJapan2008, HasanKaneRevModPhys2010}.
  Quasi-helical LLs can, for example, emerge if a magnetic field is applied to a quantum wire with  Rashba spin-orbit interaction (SOI)~\cite{StredaPRL2003}. (Quasi-)helical LLs have applications as Cooper pair splitters~\cite{SatoPRL2010} or spin filters~\cite{StredaPRL2003}, and are an essential ingredient for topological quantum wires with Majorana bound states~\cite{AliceaRepProgPhys2012}.

\begin{figure}
	\includegraphics[width=\columnwidth]{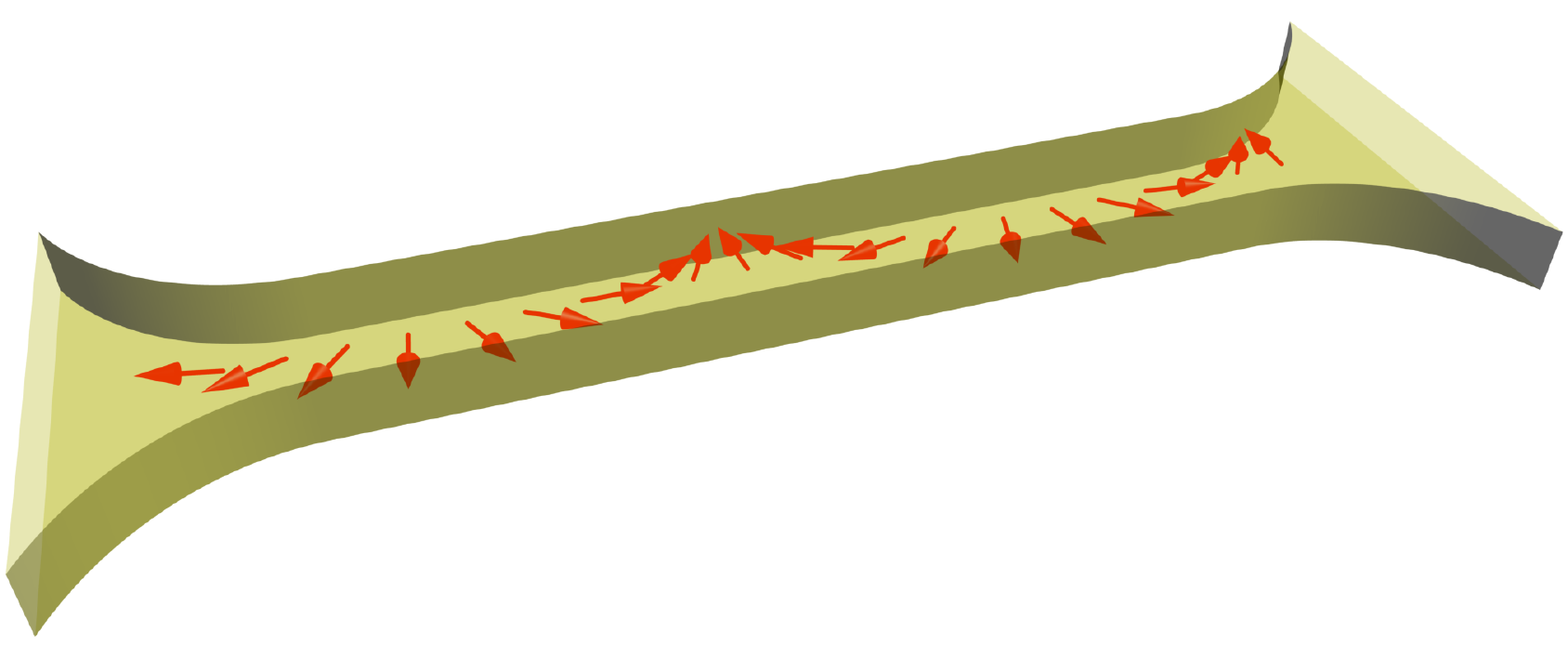}	
	\caption{(Color online) A sketch of a 1D quantum wire  with itinerant electrons
		(not shown) coupled to localized nuclear spins (red
		arrows) via hyperfine interaction. 
	A helical nuclear spin polarization
		emerges below a critical temperature \cite{BrauneckerPRB2009}.}
	\label{fig:helix}
\end{figure}

Quasi-helical LLs can also be generated by a helical magnetic field which is equivalent to the combination of a homogeneous magnetic field and Rashba SOI~\cite{BrauneckerPRB2010}. An intrinsic helical magnetic field arises as a result of hyperfine coupling between interacting electrons and nuclear spins: the Ruderman-Kittel-Kasuya-Yosida (RKKY) \cite{FrohlichProcSocLondon1940,RudermanKittelPhysRev1954, KasuyaProgTheorPhys1956, YosidaPhysRev1957} interaction diverges at momentum $2k_F$ due to electron backscattering inducing a helical order of nuclear spins~ \cite{BrauneckerPRL2009, BrauneckerPRB2009, MengPRB2013, MengEPLJB2014, HsuPRB2015}~(see Fig.~\ref{fig:helix}). This helical order reveals itself as a spatially rotating Overhauser field acting on electrons. As a result, a partial gap strongly enhanced by electron-electron interactions opens around the Fermi level. While in infinite systems this order would be suppressed by long-wavelength magnons~\cite{LossPRL2011}, the helical order still can exist in a finite-length wire~\cite{BrauneckerPRB2009, MengEPLJB2014}.

The possible experimental evidence for a nuclear spin order has been observed by Scheller~\textit{et al.}~\cite{SchellerPRL2014} by measuring temperature dependence of conductance in a cleaved edge overgrowth GaAs quantum wire. Remarkably, at low temperatures, the conductance is $G_0 \equiv e^2/h$  instead of the expected $2G_0$ for a single channel in a spin-degenerate quantum wire. At higher temperatures the conductance becomes $2G_0$ \cite{SchellerPRL2014}.  This can be explained by the lifting of electron spin degeneracy at low temperatures in the presence of a helical nuclear spin order~\cite{BrauneckerPRL2009, BrauneckerPRB2009}. Further ways to confirm the presence of the nuclear spin helix were suggested theoretically, for example, by means of nuclear magnetic resonance \cite{StanoLossPRB2014}, nuclear spin relaxation \cite{ZyuzinPRB2014}, and quantum Hall effect anisotropies~\cite{MengEPLJB2014}.

In this paper we study the temperature-dependence of the conductance in an interacting quantum wire with a helical nuclear spin order.
Although the conductance at finite temperatures in quasi-helical one-dimensional~(1D) electron systems (namely, in  wires with Rashba SOI) has been previously studied in~Ref.~\onlinecite{SchmidtPRB2014}, the main attention has been paid to weakly-interacting electrons. For zero-temperature and finite frequency conductances in strongly interacting Rashba wires see Ref. \onlinecite{Meng_FritzPRB2014}.
However, an essential ingredient for formation of the helical nuclear spin order is a strong electron backscattering, and, thus, our aim is to investigate how interactions affect the conductance of a quantum wire with a nuclear spin order at finite temperatures. 
Using a bosonization framework we describe the gapped electron modes by sine-Gordon-like solitons or kinks. These kinks are coupled to gapless excitations which leads to an  Ohmic-like friction for these kinks and thus to a temperature-dependent resistivity. 
As a result, the conductance rises from $G=G_0$ at temperatures below the critical temperature when nuclear spins are fully polarized to $G=2G_0$ at higher temperatures when the order is destroyed, featuring a relatively wide plateau in the intermediate regime, in qualitative agreement with  the experimental observation by Scheller~\textit{et~al.} \cite{SchellerPRL2014}. 
Allowing in addition for different temperatures in the  nuclear spin and electron system, the  data can be fitted by our expression for the conductance
over the entire temperature regime of the experiment.

The outline of the paper is as follows. In Sec.~\ref{sec:model}, we introduce the model of a 1D quantum wire with a helical nuclear spin order. In Sec.~\ref{sec:non-int} we discuss the electron transport in the wire disregarding electron-electron interactions both in the fermionic and bosonization frameworks. In Sec.~\ref{sec:int} we study how  interactions affect the finite-temperature conductance using one-soliton and dilute soliton gas approximations. In Sec.~\ref{sec:gap} we revise the temperature dependence of the partial gap. Finally, in Sec.~\ref{sec:discussion} we compare the theoretical results with the available experimental data. The appendices contain technical details.

\section{The model \label{sec:model}}

We consider a 1D semiconductor quantum wire of  length $L$ aligned along the $x$-axis with itinerant electrons coupled to localized nuclear spins via hyperfine coupling~(see Fig.~\ref{fig:helix}). The wire is adiabatically connected to normal Fermi liquid leads. Typically, the number of nuclear spins  $N_\perp$ in the cross-section of the wire is large, $N_\perp \gg 1$, and the hyperfine coupling constant $A$ of the material is much smaller than the Fermi energy $\varepsilon_F$, $A\ll \varepsilon_F$. We adopt then the Born-Oppenheimer approximation: since the dynamics of nuclear spins is much slower than that of electrons, the effect of nuclear spin polarization can be described as a static Overhauser field. A helical order triggered by RKKY interaction appears below a critical temperature~\cite{BrauneckerPRB2009}, the Overhauser field is spatially rotating: $\mathbf{B}(x) = B(x)\left[\mathbf{e}_x \cos (2k_F x) - \mathbf{e}_y \sin (2k_F x)\right] $, with a period determined by the Fermi momentum $k_F$. Here, for definiteness, we assume that $\mathbf{B}$ is rotating in the $xy$-plane, which need not be the case in general. The amplitude $B(x)$ is assumed to be constant inside
the wire and vanisihing in the leads (see below).
The  Hamiltonian $H$ in second quantization form for the electron subsystem and the associated Hamiltonian density $\mathcal{H}(x)$  are given by~\cite{BrauneckerPRB2009}

\begin{align}
&H = \int dx\; \mathcal{H}(x),\\
\begin{split}
&\mathcal{H}(x) =\sum\limits_s{\psi}_s(x)^\dag \left(\frac{-\partial_x^2 - k_F^2}{2m} \right){\psi}_s(x) 
\\&\hspace{80pt}+ \sum\limits_{ss'}{\psi}^\dag_s(x)\left( \bm{b} \cdot \bm{{\sigma}}_{ss'}\right) {\psi}_{s'}(x),  
\label{eqn:Hamiltonian}
\end{split}
\end{align}
where $\bm{b}(x) = g \mu_B \bm{B}(x)$, $\mu_B$ is the Bohr magneton and $g$ the electron $g$-factor, $\bm{{\sigma}}$ is a vector of Pauli matrices acting on the electron spin space, $ \psi_s(x)$ is a field operator annihilating an electron at position $x$ with spin $s=\pm $ (along the spin quantization axis $z$). Here and in the following we set $\hbar = 1$. 

The Overhauser field $B(x)$ is weak compared to the Fermi energy and  can be treated as a small perturbation. In order to describe interacting electrons with the LL model, we linearize the electron spectrum in the vicinity of Fermi points $k=\pm k_F$, and represent the fermionic fields in terms of slowly-varying left ($ L_s$) and right ($ R_s$) mover fields,
$
{\psi}_s(x) = {R}_s (x) e^{i k_F x} + {L}_s (x) e^{-i k_F x}.
$
The Hamiltonian~density given by Eq.~(\ref{eqn:Hamiltonian}) can be now rewritten as 
\begin{multline}
 \mathcal{H}(x) = v_F\sum\limits_{s}\left[  R_s^\dag (-i\partial_x)  R_s +  L_s^\dag (i\partial_x)  L_s	\right]\\+  b(x) \left[  R_\up^\dag  L_\down +  L_\down^\dag  R_\up  \right],
\label{eqn:Hamiltonian-lin}
\end{multline}
where $v_F = k_F/m$ is the Fermi velocity.

\section{Non-interacting electrons \label{sec:non-int}}
\subsection{Fermionic representation}

First, we disregard electron-electron interactions, assuming that their only role is the formation of a nuclear spin order. In this case the dynamics of the electrons can be completely described in fermionic representation. 


The Hamiltonian is block-diagonalized in the basis of the fields $R_\down$, $L_\up$ describing two gapless modes with spectrum $\varepsilon^+_{1,2}(k)=\pm v_F k$  and fields $R_\up, L_\down$ describing gapped modes with spectrum $\varepsilon^-_{1,2}(k) = \pm \sqrt{b^2 + (v_F k)^2}$ (for the moment we ignore the leads), where we assumed a constant amplitude $b>0$ of the Overhauser field inside the wire. In the following the plus sign will denote the gapless modes, and the minus sign will denote the gapped modes. The gapless modes yield a temperature-independent contribution to the conductance $G_+=G_0 = e^2/h$. In the presence of the partial gap $b$ a contribution of gapped branches to the two-terminal conductance is temperature-dependent and is given by the generalized Landauer formula~\cite{ButtikerPRB1985, Datta1997},
\begin{align}
G_{-}(V=0) = G_0 \int\limits_{-\infty}^{+\infty} d\varepsilon\; \mathcal{T}(\varepsilon) \frac{1}{4T \cosh^2\left(\varepsilon/2T\right)},
\label{eqn:Landauer}
\end{align}
where $V$ is the applied voltage difference and $\mathcal{T}(\varepsilon)$ is a transmission coefficient for the gapped modes. Here and in the following we set $k_B = 1$ 

\begin{figure}
	\includegraphics[width=\columnwidth]{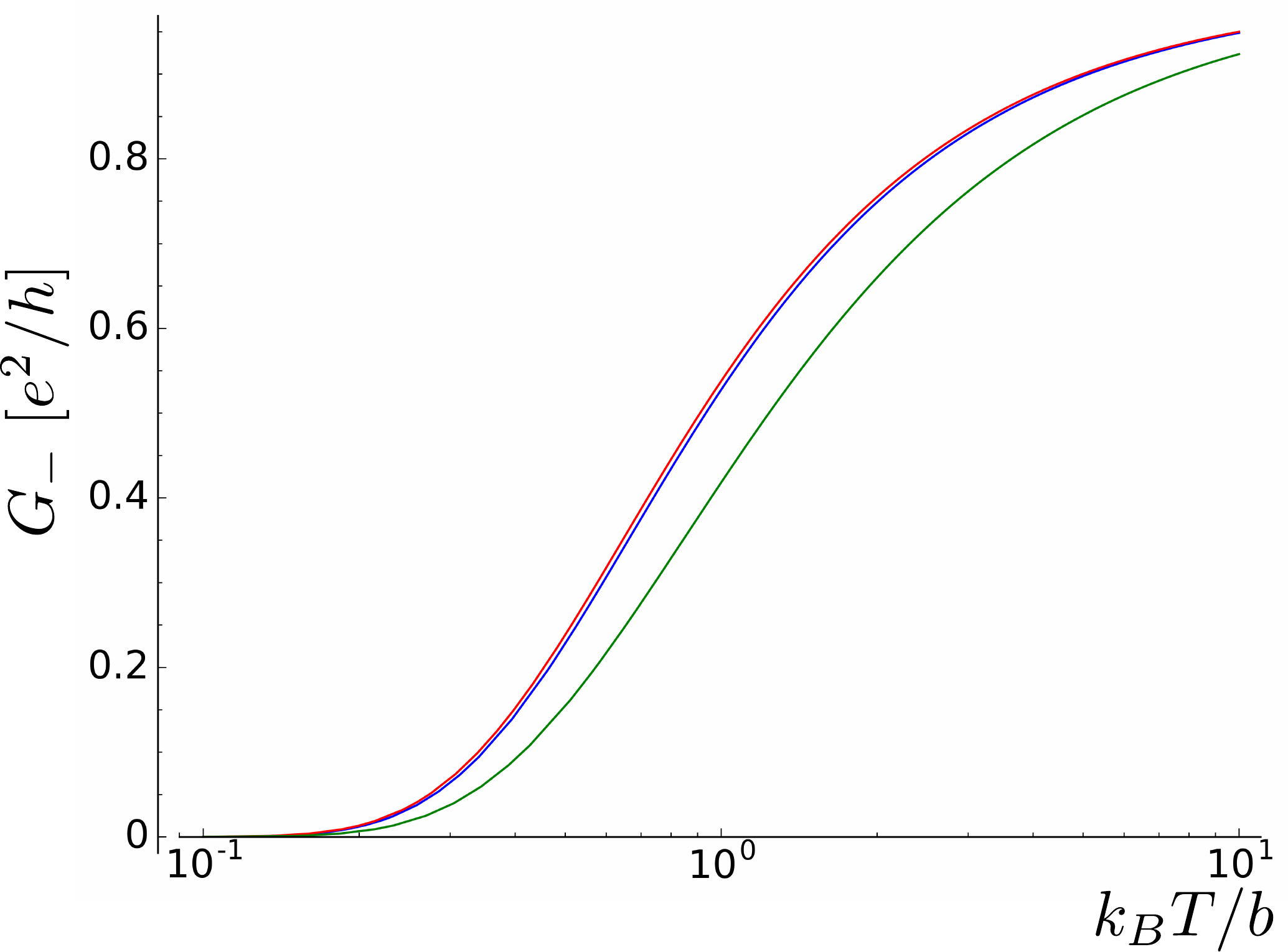} 	
	\caption{
(Color online) Conductance $G_{-} $ of the gapped mode for non-interacting electrons as function of temperature $T$ (scaled by $b=g\mu_B B$) for three models: the Overhauser field $ B(x) $  vanishes adiabatically in the leads~(red line); 
$B(x)$ 	vanishes abruptly at the contacts~(green line); smooth dependence of $B(x)$ on position $x$ given by~Eq.~(\ref{eqn:B-tanh-maintext}) with $l=l_B=\hbar v_F/b$ (blue line). The wire length $L$ is taken to be much longer than $l_B$, $L=20 l_B$.}
	\label{fig:nonint-conductance}
\end{figure}

If the wire is long enough compared to the magnetic length $l_B = \hbar v_F/b$, the tunneling current ({\it i.e.}, the contribution from energies below the gap $|\varepsilon|<b$) can be neglected, and the integration can be performed only for the energies above the gap~(see details in the Appendix~\ref{app:fermionic}). The transmission coefficient $\mathcal{T}(\varepsilon)$ and, hence, the temperature-dependent conductance itself are not universal in the sense that they depend on how the Overhauser field $B(x)$ varies close to the leads. First, we assume that the Overhauser field vanishes in the leads,  {\it i.e.}, at $x<0$, $x>L$, and then abruptly turns on to some constant finite value $B$ in the wire at $0<x<L$, $B(x) = B\Theta(x)\Theta(L-x)$, where $\Theta(x)$ is the Heaviside step-function. For energies above the gap, $|\varepsilon|>b$, we obtain (see Appendix~\ref{app:fermionic})
\begin{align}
\mathcal{T}(\varepsilon) = \frac{\varepsilon^2 - b^2}{\varepsilon^2 - b^2 \cos^2\left( \sqrt{\varepsilon^2 -b^2}L/v_F\right)}.
\end{align}
For long enough wires $L>l_B$, the transmission coefficient in~Eq.~(\ref{eqn:Landauer}) can be replaced by its averaged value $\bar{\mathcal{T}}(\varepsilon) = \sqrt{\varepsilon^2-b^2}/\varepsilon$. The contribution to the conductance $G_{-}$ by the gapped electrons as function of temperature $T$  is shown in Fig.~\ref{fig:nonint-conductance}~(green line). At low temperatures it is proportional to $\sqrt{T/b}\exp\left(-b/T\right)$.

In the opposite limiting case, the magnetic field adiabatically changes from zero in the leads to the finite value in the wire. The transmission coefficient in~Eq.~(\ref{eqn:Landauer}) can be taken equal to unity above the gap and zero below the gap $\mathcal{T}(\varepsilon) \approx \Theta(|\varepsilon|-b)$. The conductance is given in this case by 
\begin{align}
G_{-} \approx G_0\left[ 1-\tanh \left(b/2T\right) \right],
\label{eqn:conductance-smooth}
\end{align}
and is shown in~Fig.~\ref{fig:nonint-conductance}~(red line). At low temperatures the conductance is described by the activation law $G_{-} \approx 2G_0\exp\left(-b/T\right)$.

We also considered numerically an intermediate case of a smoothly varying Overhauser field $B(x)$. The coordinate dependence is modeled as follows,
\begin{equation}
B(x) = \frac{B}{2}\left(\tanh \frac{x}{l} + \tanh \frac{L-x}{l} \right).
\label{eqn:B-tanh-maintext}
\end{equation}
It turns out that if the length $l$ over which the Overhauser field varies satisfies $l\gtrsim l_B$, the numerically obtained conductance is close to the result for the ideal transmission given by Eq.~(\ref{eqn:conductance-smooth}) (see~Fig.~\ref{fig:nonint-conductance}). Thus, in the following we can assume that the transmission is ideal $\mathcal{T}(\varepsilon) = 1$, and the Overhauser field does not depend on the position $x$ (or adiabatically vanishes in the leads).

\subsection{Bosonization\label{sec:bosonization}}
It is instructive to obtain the same result for the conductance of non-interacting electrons in bosonization representation. The bosonized Hamiltonian density 
reads \cite{GiamarchiBook,VoitRepProg1995}
\begin{multline}
 \mathcal{H} =\; \frac{v_F}{2\pi}\sum\limits_{\nu=\rho,\sigma}\left[ \left( \partial_x \theta_\nu \right)^2 + \left( \partial_x \phi_\nu \right)^2  
\right]\\  
   -\frac{b}{\pi a} \cos\left[\sqrt{2}\left(\phi_\rho - \theta_\sigma \right) \right],
\end{multline}
where the conjugate bosonic fields $\phi_{\rho(\sigma)}$, $\theta_{\rho(\sigma)}$ describe the charge~(spin) sector, and $a \sim \hbar v_F/\varepsilon_F$ is a short-distance cut-off. 

For the sake of simplicity we assume that the Overhauser field $B(x)$ is position-independent (as discussed in the previous section).

It is convenient to introduce bosonic fields $\phi_{-(+)}$, $\theta_{-(+)}$ corresponding to the gapped~(gapless) branches,
\begin{align}
\phi_{\mp} &=\; \frac{\phi_\rho \mp \theta_\sigma}{\sqrt{2}},\quad
\theta_{\mp} =\; \frac{\theta_\rho \mp \phi_\sigma}{\sqrt{2}}. \label{eqn:new-vars}
\end{align}
In the new variables the Hamiltonian can be rewritten as a sum of two independent Hamiltonian densities, $\mathcal{H}=\mathcal{H}_{-} + \mathcal{H}_+$, for gapped and gapless modes,
\begin{align}
\mathcal{H}_+(x)& = \frac{v_F}{2\pi}\left\{  (\partial_x \theta_+)^2 + (\partial_x \phi_+)^2   \right\},\\
\mathcal{H}_-(x)& = \frac{v_F}{2\pi}\left\{  (\partial_x \theta_-)^2 + (\partial_x \phi_-)^2\right\}  -  \frac{b}{\pi a}\cos (2 \phi_-) . 
\end{align}
The first term $\mathcal{H}_+$ is a standard LL Hamiltonian, while  $\mathcal{H}_-$ is exactly the~sine-Gordon Hamiltonian. 

The charge current is related to the bosonic field $\phi$ by
\begin{align}
&j=j_+ + j_-,\label{eqn:current-1}\\
&j_{\pm} = \partial_t \phi_\pm/\pi,\label{eqn:current-2}
\end{align}
 and according to the Maslov-Stone approach~\cite{MaslovStonePRB1995}, the full conductance $G$ can be extracted from the retarded Green functions $ \mathcal{G}_{\pm}^R(\omega)$ via the Kubo formula,
\begin{align}
G = \frac{e^2}{\pi^2 }\lim\limits_{\omega\to 0} \omega\left[ \mathcal{G}_+^R(\omega)+\mathcal{G}_-^R(\omega)\right].
\label{eqn:Kubo-retarded}
\end{align}
The definitions of the Green functions are given in~Appendix~\ref{app:Green-def}. We will use also the Matsubara version of Eq.~(\ref{eqn:Kubo-retarded}), which is given by analytical continuation,
\begin{align}
G = \left.\frac{e^2}{\pi^2}\bar{\omega}\left[ \mathcal{G}_+^M(x,x,\bar{\omega})+\mathcal{G}_-^M(x,x,\bar{\omega})\right]\right\rvert_{i\bar{\omega}\to \omega+i0,\;\omega\to 0}.
\label{eqn:Kubo-Matsubara}
\end{align}

Currents corresponding to the gapped and gapless modes commute with each other, so their contribution can be calculated independently, {\it i.e.}, $G=G_{+} +G_{-}$.  
The gapless fields yield the conductance $G_{+} = G_0$, while the gapped ones give rise to $G_{-}$. From now on we focus on the gapped modes.

The Hilbert space for the sine-Gordon model consists of the vacuum sector (with vacuum state $|0\rangle$ and fluctuations around it) and the sectors with different number of kinks, antikinks, and the bound states~\cite{Rajaraman1982,GoldstoneJackiwPRD1975}. The one-kink sector is orthogonal to the vacuum sector and consists of the following. (i) kink-particle states $\left| P \right\rangle$ with mass $M$, momentum $P$, and energy $\varepsilon(P) = v_F \sqrt{P^2  + M^2 v_F^2}$; (ii) scattering states $\left| P, k_1,k_2,\dots, k_n \right\rangle$ of the kink-particle and $n$ `mesons' (fluctuations around the kink) with asymptotic momenta $P,k_1, \dots, k_n$.

The sine-Gordon kink-particles are important for describing the transport for the gapped modes since these kinks carry electric current and their topological charge corresponds to the electric charge.

The mass of the kink~\cite{ZamolodchikovIntJModPhysA1995} in the non-interacting case is related to the Overhauser field energy by $Mv_F^2 = b$ which is in agreement with the fermionic picture (the one-kink sector corresponds to electron states above the gap). In this section we will restrict our study to the one-kink sector, disregarding multi-kink states, which can be justified at low temperatures $T \ll b$.

For the sake of simplicity, in this section we will also disregard scattering states $\left| P, k_1,k_2,\dots, k_n \right\rangle$. Since in the non-interacting case mesons have a finite mass, this can be justified in the same limit of low temperatures. The effect of mesons on the conductance will be discussed in Sec.~\ref{sec:one-kink}, when we consider the more general interacting case.

\subsubsection{Vacuum sector}

In order to consider the system in the vicinity of the vacuum state we replace the cosine term in the Hamiltonian with a quadratic one, using a self-consistent harmonic approximation~\cite{GiamarchiBook},

\begin{align}
&\frac{b}{\pi a}\cos (2 \phi_-) \to \mathrm{const} -\frac{\Delta^2}{2\pi v_F} \phi_-^2,\\ 
&\Delta^2= \frac{2 b v_F}{a}e^{-2\langle \phi_-^2 \rangle}.
\end{align}

Following the Maslov-Stone approach~\cite{MaslovStonePRB1995} we assume infinitesimal dissipation in the leads, so that the Matsubara Green function vanishes far away from the contacts  $\mathcal{G}^M_{\bar{\omega}}(x=\mp \infty,x') = 0$. The Matsubara Green function $\mathcal{G}^M_{\bar{\omega}}(x,x')$ in the wire can be calculated straightforwardly,
\begin{align}
\mathcal{G}^M_{\bar{\omega}}(x,x') = - \frac{\pi}{2\sqrt{\bar{\omega}^2 + \Delta^2}}\exp\left\{ -\frac{\sqrt{\bar{\omega}^2 + \Delta^2 }|x-x'|}{v_F} \right\}.
\end{align}
Since the Matsubara Green function is finite in the low-frequency limit $\bar{\omega} \to 0$, the Kubo formula, see Eq.~(\ref{eqn:Kubo-Matsubara}), yields zero conductance contribution from the gapped modes.

\subsubsection{One-kink sector}
Here we will follow a method for calculating low-temperature correlation functions described in~Ref.~\onlinecite{Essler2005}. First, we expand out the thermal trace
\begin{multline}
\left\langle \phi_-(\tau) \phi_-(0) \right\rangle =\; \frac{1}{\mathcal{Z}}\mathrm{Tr}\left\{ e^{-\beta H} \phi_-(\tau) \phi_-(0) \right\} \\
=\; \frac{1}{\mathcal{Z}}\left\{\left\langle 0\middle| \phi_-(\tau) \phi_-(0) \middle|0\right\rangle + \sum\limits_Pe^{-\beta \varepsilon(P)}\left\langle P\middle| \phi_-(\tau) \phi_-(0) \middle|P\right\rangle \right.\\+\left. \sum\limits_{\bar{P}}e^{-\beta \varepsilon(\bar{P})}\left\langle \bar{P}\middle| \phi_-(\tau) \phi_-(0) \middle|\bar{P}\right\rangle + \cdots \right\},
\label{eqn:trace}
\end{multline}
where $\beta = T^{-1}$ is the inverse temperature, $\left| P\right\rangle$ and $\left| \bar{P}\right\rangle$ are kink and anti-kink states with momentum $K$ and $\bar{K}$ and energies $\varepsilon(K) = \sqrt{v_F^2 K^2+b^2}$ and $\varepsilon(\bar{K})=\sqrt{v_F^2\bar{K}^2+b^2}$, respectively, $\mathcal{Z}$ stands for the partition function. We will restrict ourselves to the low temperature limit $T\ll b$ and disregard higher-order sine-Gordon solitons. The first (``vacuum'') term in~Eq.~(\ref{eqn:trace}) yields zero contribution to the current as discussed in the previous section.

The two-point correlator in the kink state can be expressed via the matrix elements of $\phi$ as
\begin{multline}
\left\langle P\middle| \phi_-(x,\tau) \phi_-(0)\middle| P \right\rangle = 
\sum\limits_{P'} \left\langle P \middle| \phi_-(\tau)\middle| P' \right\rangle \left\langle P'\middle| \phi_-(0)\middle| P \right\rangle\\=
\sum\limits_{P'}e^{- \left[\varepsilon(P)-\varepsilon(P')\right]\tau} e^{- i(P-P')x }
\left|\left\langle P\middle|\phi_-\middle|P' \right\rangle\right|^2.
\label{eqn:correlator}
\end{multline}

The matrix elements can be calculated in the quasiclassical limit~\cite{GoldstoneJackiwPRD1975, MussardoNuclPhysB2003} ($P v_F \ll b$) as the Fourier transformation of the static kink solution $\phi_{K} = 2\arctan \exp\left(x/\delta_0\right)$,
\begin{multline}
\left\langle P \middle| \phi_- \middle| P' \right\rangle = b \int dx\; e^{i(P-P') x} \phi_{K}(x) \\\approx
\frac{\pi i }{(P - P') \cosh [\gamma (P-P')]} + \pi^2\delta (P-P').
\end{multline}
where $\delta_0=\sqrt{\hbar v_F a/2b}$ is the width of the kink, and $\gamma= \pi b \delta_0/\hbar v_F$.
The latter term with the delta-function will result in a divergent but time-independent contribution to the correlation functions, and therefore will vanish after taking the time derivative.

Now we can perform the integration over $P$, $P'$ in~Eqs.~(\ref{eqn:trace})--(\ref{eqn:correlator}) and obtain a contribution to the Matsubara Green function from the kink-sector,
\begin{align}
\left\langle \phi_-(\bar{\omega}, x)\phi_-(-\bar{\omega},x) \right\rangle_{k} = \frac{1}{2\bar{\omega}}e^{-b/T}.
\end{align}

Similarly, there will be the equal contribution from the anti-kink sector. The total conductance by the gapped modes can be calculated by the Kubo formula~(\ref{eqn:Kubo-Matsubara}),
$$G_{-} = 2G_0 e^{-b/T}, $$
which agrees with the low-temperature expansion of Eq.~(\ref{eqn:conductance-smooth}) obtained in the fermionic representation.

\section{Interacting wire \label{sec:int}}

In the general case with interactions the LL is described by charge and spin interaction parameters $K_\rho$, $K_\sigma$. In the following we put $K_\sigma=1$. The LL parameter $K_\rho$ varies from zero for strong (unscreened) electron repulsion to $K_\rho=1$ for non-interacting electrons. In order to treat the leads correctly, we assume similarly to Ref.~\onlinecite{MaslovStonePRB1995} that the interaction parameter depends on the coordinate $x$, and there is no interaction in the leads $K_\rho(x) =1$ at $x<0$, $x>L$. 
The Hamiltonian density describing interacting electrons in the Overhauser field $b$ is given by~\cite{BrauneckerPRB2009,MengEPLJB2014}
\begin{multline}
\mathcal{H} = \frac{v_F}{2\pi}\sum\limits_{\nu=\rho,\sigma}\left\{ \left(\partial_x \theta_{\nu}\right)^2 + \frac{1}{K_\nu^2}\left(\partial_x\phi_{\nu} \right)^2  \right\} \\- \frac{b}{\pi a} \cos\left[\sqrt{2}\left( \phi_\rho -\theta_\sigma \right)  \right] 
\end{multline}

In terms of the gapped and gapless fields $\phi_-$, $\phi_+$ and dual fields $\theta_-$, $\theta_+$ defined by~Eq.~(\ref{eqn:new-vars}) the Hamiltonian density takes the form
\begin{multline}
\mathcal{H} = \left.\frac{v_F}{2\pi}\middle\{ (\partial_x \theta_-)^2 +(\partial_x \theta_+)^2 \right.\\ \left. + \frac{K_\rho^{-2}+1}{2}\left[\left(\partial_x \phi_- \right)^2 + \left(\partial_x \phi_+ \right)^2 \right]\right. \\\left. + \frac{K_\rho^2 -1}{2} 2\left(\partial_x \phi_1 \partial_x \phi_2 \right) \right\} 
- \frac{b}{\pi a} \cos\left(2\phi_-\right) .
\end{multline}

After integrating out dual fields $\theta_+$, $\theta_-$ the Euclidean action $S_E = S_++S_-+S_{+-}$ consists of the sine-Gordon action $S_-$ describing gapped modes,

\begin{multline}
S_- =\; \frac{1}{\pi v_F}\int dxd\tau\; \left\{ \frac{\left(\partial_\tau \phi_- \right)^2}{2} + v_+^2(x) \frac{\left(\partial_x \phi_- \right)^2}{2} 
\right.\\\left. - \frac{b v_F}{ a}\cos (2\phi_-)\right\},
\label{eqn:interior-interaction}
\end{multline}
a standard LL action $S_+$ describing gapless modes,
\begin{align}
S_+ &=\; \frac{1}{\pi v_F}\int dxd\tau\; \left\{ \frac{\left(\partial_\tau \phi_+ \right)^2}{2} + v_+^2(x) \frac{\left(\partial_x \phi_+ \right)^2}{2}\right\},
\end{align}
and in the general interacting case there also appears a coupling between the gapped and the gapless modes,
\begin{align}
S_{+-} &=\; \frac{1}{\pi v_F}\int dxd\tau\; v_-^2(x) \partial_x\phi_+ \, \partial_x \phi_- \, ,
\end{align}
where $v_{\pm}^2 = v_F^2\left(K_\rho^{-2}(x)\pm 1\right)/2$.  Note that the cross term $S_{+-}$ vanishes in a non-interacting system ($K_\rho=1$). 

The cosine term in~Eq.~(\ref{eqn:interior-interaction}) is relevant in the renormalization-group (RG) sense and leads to the gap $\Delta$ in the spectrum. The renormalized gap is given by~\cite{BrauneckerPRB2009}
\begin{align}
\Delta = b \left( \frac{l_\xi}{a} \right)^{(1-K_\rho)/2},
\label{eqn:gap}
\end{align}
with the correlation length of the gapped modes $l_\xi = \min\left\{L, \hbar v_F /T, \hbar v_F/\Delta   \right\}$.

The Euler-Lagrange equations for the action $S_E$ read
\begin{align}
&\partial_\tau^2 \phi_- + \partial_x[v_+^2(x) \partial_x \phi_-]  + \partial_x[v_-^2(x)\partial_x\phi_+] = 2\frac{b v_F}{ a}\sin 2\phi_-,\\
&\partial_\tau^2 \phi_+ + \partial_x[v_+^2(x) \partial_x \phi_+]  + \partial_x[v_-^2(x)\partial_x\phi_-]=0.
\label{eqn:Euler-Lagrange-ext}
\end{align}

A static solution can be found by taking $\partial_\tau \phi_{\pm}=0$ and expressing $\partial_x \phi_+$ from~Eq.~(\ref{eqn:Euler-Lagrange-ext}): $\partial_x \phi_+ = -v_-^2\partial_x \phi_-/v_+^2$. The resulting equation for $\phi_-$ resembles the sine-Gordon equation,
\begin{align}
\partial_x \left[ c^2 \partial_x \phi_- \right] = 2\frac{b v_F}{a}\sin 2 \phi_-,
\end{align}
with the effective ``speed of light'' $c^2 = (v_+^4-v_-^4)/v_+^2$ which takes values from $c(K_\rho=1)=v_F$ for non-interacting electrons to $c(K_\rho \to 0)= \sqrt{2} v_F$ for strongly-interacting electrons.

The vacuum classical static solution is trivial $\phi_+^{0} = \phi_-^{0} = 0$.  If the Overhauser field $b$ and the interaction parameter $K_\rho$ depend on the coordinate adiabatically, the classical static solution inside the wire for an (anti)kink with  center at $x=\xi$ is given by
\begin{align}
\phi_-^{k(a)}(x) = 2\arctan \exp \left\{\pm\int\limits_{\xi}^x \frac{dx'}{\delta_0(x')} \right\}.
\label{eqn:static-kink}
\end{align}
The plus sign corresponds to a kink solution, while the minus sign corresponds to an antikink solution, and $\delta_0(x) = c(x)\sqrt{\hbar a/2b(x)v_F}$ is the soliton width.

 The Matsubara current-current correlator can be expressed by using functional integration over the fluctuations in the vicinity of the classical static solutions

\begin{multline}
\left\langle j(\tau) j(0)\right\rangle = \frac{1}{\mathcal{Z}}\left\{ \int \mathcal{D}\delta\check{\varphi}\; j(\tau)j(0)e^{-S_E[\check{\phi}^{0}+\delta \check\varphi]} \right.\\ +\left. \int \mathcal{D}\delta\check{\varphi}\; j(\tau)j(0)e^{-S_E[\check{\phi}^{k}+\delta \check\varphi]}+\dots\right\},
\label{eqn:functional-formulation}
\end{multline}
\begin{align*}
\mathcal{Z} = \int \mathcal{D}\delta\check{\varphi}\; e^{-S_E[\check{\phi}^{0}+\delta \check\varphi]} + \int \mathcal{D}\delta\check{\varphi}\; e^{-S_E[\check{\phi}^{k}+\delta \check\varphi]}+\dots
\end{align*}
Here we use a short-hand notation $\check{\phi} = \begin{pmatrix}\phi_- ,& \phi_+ \end{pmatrix}^T$. In the vicinity of a classical one-kink solution $\check{\phi}^{k}$, we expand the fields as a sum of classical solution and fluctuations around it,
\begin{align}
\check{\phi}(x,\tau) = \check{\phi}^{k}(x,\xi) + \delta\check{\varphi}(x-\xi,\tau),
\label{eqn:expansion}
\end{align}
and treat the center of the kink as a dynamical variable $\xi(\tau)$. However, this representation is redundant: shifts of both the collective coordinate $\xi$ and the Goldstone zero-mode $\delta \check{\varphi} \propto \partial_x \check{\phi}^{k}$ describe a translation. In order to avoid double counting we have to impose the following constraint. The integration is performed only over the fluctuations orthogonal to the zero-mode $\int dx\; \delta \check{\varphi}(x,\tau)\partial_x \check{\phi}^{k}(x) = 0$. This can be done by the Faddeev--Popov technique~\cite{GervaisSakitaPRD1975,Sakita1985}. The integrals over the fluctuations near static kink solutions in~Eq.~(\ref{eqn:functional-formulation}) have to be rewritten as
\begin{multline}
\int \mathcal{D}\delta\check{\varphi}\; e^{-S_E[\check{\phi}^{k}+\delta \check\varphi]} \dots \to\\ \int \mathcal{D}\delta \check{\varphi}\mathcal{D}\xi\;\delta\left(Q[\xi]\right)
 \det (\frac{\delta Q}{\delta \xi}) e^{-S_E[\check{\phi}^{k}+\delta \check\varphi]}\dots, 
\end{multline}
with the Faddeev-Popov functional $Q[\xi] = \int dx\; \check{\phi}(x,\tau)\partial_x \check{\phi}^{k}(x,\xi)$.

Using expansion~(\ref{eqn:expansion}) and relating the current to bosonic fields by~Eqs.~(\ref{eqn:current-1})--(\ref{eqn:current-2})
 we represent the current-current correlator as an (anti)kink-particle contribution and ``background fluctuations'' contribution
\begin{align}
\begin{split}
\langle j(\tau) j(0) \rangle &=\; \langle j(\tau) j(0) \rangle_{k} + \langle j(\tau) j(0) \rangle_{a} \\&+
\langle \delta j(\tau) \delta j(0) \rangle \label{eqn:current-expansion1},
\end{split}\\
\langle j_\pm(\tau) j_\pm(0) \rangle_{k(a)} &=\; -\frac{1}{\pi^2}\partial_\tau^2 \langle \check{\phi}^{k(a) }_\pm(x,\tau)\check{\phi}_\pm^{k(a)}(x,0)  \rangle \label{eqn:current-expansion2},\\
\langle \delta j_\pm(\tau) \delta j_\pm(0) \rangle &=\; -\frac{1}{\pi^2}\partial_\tau^2 \langle \delta\check{\phi}_\pm(x,\tau) \delta\check{\phi}_\pm(x,0)  \rangle \label{eqn:current-expansion3}.
\end{align}
The electric current does not depend on the coordinate $x$. However, the calculations are easier if we calculate the correlators in the leads, taking $x<0$ in~Eqs.~(\ref{eqn:current-expansion2})--(\ref{eqn:current-expansion3}), where the stationary classical solution for gapless modes turns to zero. In this case it is sufficient to calculate the correlators for the gapped field $\phi_-$ in~Eqs.~(\ref{eqn:current-expansion1})--(\ref{eqn:current-expansion3}).

\subsection{Current carried by a single kink-particle \label{sec:one-kink}}

At finite but low temperatures $T\lesssim \Delta$ a kink with the rest energy $\Delta$ and the mass $M=\Delta/c^2$ can be activated. The kink can propagate inside the wire carrying electric charge and interacting with the environment consisting of gapless and gapped modes of background fluctuations (see Appendix~\ref{app:eff-action} for details). The spectrum of fluctuation modes is given by
\begin{align}
\left(\omega^{\pm}_q\right)^2  
 = \frac{(c/\delta_0)^2+2 q^2 v_+^2\pm\sqrt{(c/\delta_0)^4+4 q^4 v_-^4}}{2}.
\end{align}
The plus sign corresponds to the gapped mode $\omega_q^{+}\approx \sqrt{(c/\delta_0)^2 +q^2 v_+^2}$, while the minus sign corresponds to a gapless acoustic mode: $\omega_q^- \approx v_+|q|$ at $q\delta_0 \ll 1$, $\omega_q^- \approx \sqrt{v_+^2-v_-^2}q$ at $q\delta_0 \gg 1$.

While the gapped mode leads to renormalization of the kink mass, which is described by~Eq.~(\ref{eqn:gap}), the coupling to the gapless mode causes an effective friction: the kink dissipates energy interacting with the gapless mesons. The mechanism resembles Caldeira-Leggett type dissipation~\cite{CaldeiraLeggett1983}, damping of Bloch walls in quasi-1D ferromagnets caused by interaction with spin waves~\cite{BraunPRB1996}. It is also resembles a mechanism of dissipation due to scattering of spinons in Wigner crystals~\cite{MatveevWignerCrystals}. However, in contrast to spinons, kinks carry electric current, and the resulting temperature dependence of conductance is different.

In order to calculate a contribution to the conductance due to the motion of kinks we integrate out fluctuations $\delta \check{\varphi}$ and obtain an effective low-energy Euclidean action for the collective coordinate $\xi$ (see Appendix~\ref{app:eff-action} for details of derivation),
\begin{align} 
S_E^{eff}[\xi] = \frac{\Delta}{T} + T\sum\limits_{\bar \omega}\; \xi\left(\bar{\omega}\right)\left\{\frac{M \bar{\omega}^2 }{2} + \frac{M}{2}\eta |\bar{\omega}|
\right\}\xi\left(-\bar{\omega}\right).
\end{align}
The summation over Matsubara frequencies $\bar{\omega}$ is performed. The first term is responsible for the activation law exponent; the second term describes the free motion of the kink, while the third term corresponds to an Ohmic-like friction caused by the interaction between a kink and the gapless fluctuation modes. The friction coefficient is given by
\begin{align}
\eta &= \; 4\frac{T}{\Delta} \frac{c^2}{\delta_0}\frac{v_-^2}{v_+^3}.  
\end{align}

The Matsubara Green function for the collective coordinate $\mathcal{D}^M(\bar{\omega})$ in the limiting case $L\to\infty$ reads

	\begin{align}
	\mathcal{D}^{M}(\bar{\omega}) &=\; \frac{1}{ M\left( \bar{\omega}^2 + \eta |\bar{\omega}|\right)}.
	\end{align}

In order to avoid subtleties arising from proper analytic continuation in Matsubara technique, it is convenient use the Keldysh path-integral approach. The retarded~(advanced) Green function $\mathcal{D}^{R(A)}$ for the collective coordinate $\xi$ can be extracted from the Matsubara Green function by analytic continuation,
\begin{align}
\mathcal{D}^{R(A)}(\omega) &=\; \frac{1}{ M\left(\omega \pm i0 \right)\left(\omega \pm i\eta \right)},\\ 
\mathcal{D}^{R}(\tau) &=\; \frac{1}{M} \frac{1- e^{-\eta \tau}}{\eta}\Theta(\tau).
\end{align}

The Keldysh Green function can be obtained using the fluctuation-dissipation theorem
	\begin{multline}
	\mathcal{D}^K(\omega) = \left[\mathcal{D}^R(\omega)-\mathcal{D}^A(\omega)\right]\coth \frac{\omega}{2T} \\= -\frac{2i}{M \omega}\frac{\eta}{\eta^2+\omega^2}\coth \frac{\omega}{2T},
	\end{multline}
	\begin{align}
		\mathcal{D}^K(\tau)-\mathcal{D}^K(0) \approx \frac{2i}{\pi M} T\tau^2 \arctan \frac{1}{\eta \tau}.
	\end{align}
	In the absence of the friction the Keldysh Green function reads
	\begin{align}
	\mathcal{D}^K(\tau)-\mathcal{D}^K(0) = \frac{i}{M}T\tau^2.
	\end{align}
	
	The Green function for the bosonic fields $\check{\phi}$ can be obtained by Keldysh functional integration 
	\begin{align}
	\begin{split}
	i\mathcal{G}_{\alpha\beta}(x,x,t,t') &=\; \langle \mathcal{T}_K \phi_-(r,t_\alpha) \phi_-(r',t'_\beta) \rangle \\&=\; \int \mathcal{D}\xi\; \phi^{k}\left(x,\xi(t)\right)\phi^{k}\left(x,\xi(t')\right) \exp\left[iS[\xi]\right]\\
	&=\;
	\int \mathcal{D}\xi\;\frac{dq\;dq'}{(2\pi)^2} \phi^{k}(q)\phi^{k}(q') e^{i q x+i q' x}\\&\times\; \exp\left[-i q \xi(t) -i q' \xi(t') \right]\exp\left(iS[\xi]\right),
	\end{split}
	\label{eqn:Green-path-integrals}
	\end{align}	
	where the indices $\alpha$, $\beta$ indicate whether the times $t$, $t'$ are taken on the upper or on the lower branch of the Schwinger-Keldysh contour, and $\phi^{k}(q)$ refers to the Fourier transformation of the static kink solution~(\ref{eqn:static-kink}).
	
	The functional integration is performed in the Appendix~\ref{app:path-integrations}. Finally, we express the retarded Green for bosonic fields $\phi_-$ at coincident coordinates inside the wire $x=x'$ in terms of the Green functions of the collective coordinate $\mathcal{D}^{R,A,K}$,
	\begin{multline}
	\mathcal{G}^R(x,x,\tau) = \mathcal{G}_{++}(x,x,\tau) - \mathcal{G}_{+-}(x,x,\tau) = 4\pi \Theta(\tau)\sqrt{\frac{M}{\beta}}\\\times \dfrac{\mathcal{D}^R(\tau)-\mathcal{D}^R(0)}{\sqrt{i(\mathcal{D}^K(0)- \mathcal{D}^K(\tau))}}\dfrac{\sqrt{2}}{\sqrt{\sqrt{1-\dfrac{\left(\mathcal{D}^R(\tau)-\mathcal{D}^R(0)\right)^2}{(\mathcal{D}^K(\tau)-\mathcal{D}^K(0))^2}}+1}}.
	\label{eqn:retarded}
	\end{multline}
	
		The conductance is related to low-frequency current--current correlator by~Eq.~(\ref{eqn:Kubo-retarded}), and, hence, can be extracted from the retarded Green function in time-representation at large times,
	\begin{align}
		G_{k} = e^{-\Delta/T}\lim\limits_{\tau\to+\infty} \mathcal{G}^R(\tau).
		\label{eqn:G-time-limit}
	\end{align}

The activation law exponent arises due to the rest energy $\Delta$ of the kink.
	First consider the important limiting case $\eta = 0$, $L\to \infty$. In the absence of friction, the Green functions for collective coordinate grow infinite at $\tau \to +\infty$,		
	\begin{align}
&	\mathcal{D}^{R}(\tau)\vert_{\eta \to 0} =\; \frac{\tau}{M}\Theta(\tau),\\		
&		\mathcal{D}^K(\tau)-\mathcal{D}^K(0)\vert_{\eta \to 0} =\; \frac{i}{M}T\tau^2.
	\end{align}
The kink and equal anti-kink contributions to the conductance are obtained straightforwardly from~Eq.~(\ref{eqn:G-time-limit}),	
\begin{align}
{G}_{k}={G}_{a} = G_0 e^{-\Delta/T}.
\end{align}
Thus, we see that in the absence of the friction the only effect of interactions is the renormalization of the gap $\Delta$.

The situation differs in the general case $\eta > 0$. Now the retarded Green function for the collective coordinate $D^R$ is finite at infinite times, $\mathcal{D}^R(\tau\to\infty) = \dfrac{1}{M\eta}$, but the Keldysh Green function (in the limit of infinite length $L\to\infty$) is still infinite, $\mathcal{D}^K(\tau\to \infty) \propto \tau^2$. Therefore, Eq.~(\ref{eqn:G-time-limit}) yields zero conductance.

This can be easily understood, since the Ohmic-like friction causes an internal resistivity, and we may expect that at large $L> c/\eta$ the total conductance will drop to zero with the increase of the length $L$. 

The result for finite but large wire length $L$ can be easily estimated. Since the collective coordinate is bounded inside the wire $0<\xi<L$, the Keldysh and retarded Green functions must be bounded as well, $|\mathcal{D}^K| < L^2$, $|\mathcal{D}^R| < L^2$. Therefore, we assume that the Green functions grow until they reach their asymptotic value of order of $L^2$. This gives a cut-off parameter at large times $\tau_\infty = \min\{\tau_R, \tau_K\}$ with
\begin{align}
\tau_R =  \dfrac{\Delta L^2}{\hbar c^2},\,\,\,\, \tau_K = \dfrac{L}{c}\sqrt{\dfrac{\Delta}{T}}.
\end{align}
If the cut-off time $\tau_\infty$ and the friction $\eta$ are large enough $\eta \tau_\infty \gg 1$ (this occurs at low temperatures, when the gap is large in comparison to $\hbar c/L$), the conductance is suppressed by friction,
\begin{align}
G_{k} = G_0 \sqrt{\frac{T}{\Delta}}\frac{c}{\eta(T) L}e^{-\Delta/T},\quad \eta \tau_\infty \gg 1.
\end{align}

In the opposite limit $\eta \tau_\infty \ll 1$, the friction becomes insignificant, and the conductance is the same as in the non-interacting case.

The crossover between these regimes can be roughly described by taking the limit at finite $\tau\to\tau_\infty$ instead of $\tau\to \infty$ in~Eq.~(\ref{eqn:G-time-limit}),
\begin{align}
G_{k} = 2G_0 \sqrt{T M} \frac{\dfrac{1-e^{-\eta \tau_\infty}}{M \eta} }{\sqrt{\frac{2}{\pi M} T \tau_\infty^2 \arctan \frac{1}{\eta \tau_\infty}}} e^{-\Delta/T}.
\label{eqn:one-soliton-result}
\end{align}

\subsection{Background fluctuations}
\subsubsection{Vacuum fluctuations}
First, we calculate the Matsubara Green functions for fluctuations around the vacuum solution. In order to do this we insert a point source $\left(\mathscr{J}_- \delta \varphi_- + \mathscr{J}_+ \delta \varphi_+\right)  \delta(x-x_0)$ into the action, then the solution of the Euler-Lagrange equations can be represented as $\delta \varphi_\alpha(x) = \mathcal{G}^M_{\alpha\beta}(x,x_0) \mathscr{J}_\beta $. The Euler-Lagrange equations for small fluctuations $\delta \check{\varphi}$ are given by
\begin{multline}
\bar{\omega}^2 \delta \varphi_- - \partial_x \left( v_+^2 \partial_x \delta\varphi_- + v_-\partial_x \delta\varphi_+  \right)
+W^2(x) \delta\varphi_- \\ = \pi v_F \mathscr{J}_- \delta(x-x'),
\label{eqn:Euler-Lagrange-fluc1}
\end{multline}
\begin{multline}
\bar{\omega}^2 \delta \varphi_+ - \partial_x \left( v_+^2 \partial_x \delta\varphi_+ + v_-\partial_x \delta\varphi_-  \right)
 \\  = \pi v_F \mathscr{J}_+ \delta(x-x'),
\label{eqn:Euler-Lagrange-fluc2} 
\end{multline}
where the potential $W^2(x) = c^2/\delta_0^2(x)$ varies adiabatically, and following~Ref.~\onlinecite{MaslovStonePRB1995} we assume that the interaction parameter $K_\rho(x)=1$, $v_+(x) = v_F$, $v_-(x)=0$  in the leads. Since the current in the wire does not depend on the coordinate, the point $x'$ can be chosen arbitrary. We take $x' < 0$ for the sake of simplicity.

Solving Eqs.~(\ref{eqn:Euler-Lagrange-fluc1})--(\ref{eqn:Euler-Lagrange-fluc2}) in the limit $\bar{\omega}\to 0$, 
similarly to~Ref.~\onlinecite{MaslovStonePRB1995}, we obtain 
\begin{align}
&\delta \varphi_- = O(\bar{\omega}^0)\mathscr{J}_- + O(\bar{\omega}^0)\mathscr{J}_+,\\
&\delta \varphi_+ = O(\bar{\omega}^0)\mathscr{J}_- + \left(\frac{\pi}{2\bar{\omega}} + O(\bar{\omega}^0) \right) \mathscr{J}_+.  
\end{align} 
Therefore for the Matsubara Green functions in the leads,  in the limit $\bar{\omega}\to 0$, we obtain
\begin{align*}
&\left\langle \delta \varphi_-(\tau) \delta \varphi_-(0) \right\rangle_0 = O(\bar{\omega}^0), \quad \left\langle \delta \varphi_+(\tau) \delta \varphi_-(0) \right\rangle_0 = O(\bar{\omega}^0),\\
&\left\langle \delta \varphi_+(\tau) \delta \varphi_+(0) \right\rangle_0 = \frac{\pi}{2\bar{\omega}} +O(\bar{\omega}^0).
\end{align*}
\subsubsection{Fluctuations around the kink solution}
Now we expand the fields around the  static soliton solution $\check{\phi} = \check{\phi}^k(x,x_0) + \delta \check{\varphi}$. We again insert 
a point source $\left(\mathscr{J}_- \delta \varphi_- + \mathscr{J}_+ \delta \varphi_+\right)  \delta(x-x_0)$ into the action and solve the Euler-Langrange equations which have the form of Eqs.~(\ref{eqn:Euler-Lagrange-fluc1})--(\ref{eqn:Euler-Lagrange-fluc2}), but with the potential $W^2(x) = \left(c/\delta_0(x)\right)^2 \left(1 - 2\mathrm{sech}^2\left(x/\delta_0\right) \right)$.

Similarly to Ref.~\onlinecite{MaslovStonePRB1995}, in the limit $\bar{\omega}\to 0$ and if $x<0$, $x'<0$ the result does not depend on the specific form of $W(x)$,
\begin{align}
&\delta \varphi_- = O(\bar{\omega}^0)\mathscr{J}_- + O(\bar{\omega}^0)\mathscr{J}_+,\\
&\delta \varphi_+ = O(\bar{\omega}^0)\mathscr{J}_- + \left(\frac{\pi}{2\bar{\omega}} + O(\bar{\omega}^0) \right) \mathscr{J}_+.  
\end{align} 
For the Matsubara Green functions in the leads, in the limit $\bar{\omega}\to 0$, we obtain 
\begin{align*}
&\left\langle \delta \varphi_-(\tau) \delta \varphi_-(0) \right\rangle_k = O(\bar{\omega}^0), \quad \left\langle \delta \varphi_+(\tau) \delta \varphi_-(0) \right\rangle_k = O(\bar{\omega}^0),\\
&\left\langle \delta \varphi_+(\tau) \delta \varphi_+(0) \right\rangle_k = \frac{\pi}{2\bar{\omega}} +O(\bar{\omega}^0).
\end{align*}
The correlators for the fluctuations around the antikink solution are the same.

Finally, for the contribution to the conductance from background fluctuations we obtain
\begin{align}
G = \frac{e^2}{\pi^2}\lim\limits_{\bar{\omega}\to 0}\bar{\omega} \dfrac{ \dfrac{\pi}{2\bar{\omega}}  +  2 \dfrac{\pi}{2\bar{\omega}} e^{-\frac{\Delta}{T}} +\dots}{1+2e^{-\frac{\Delta}{T}} + \dots} +O(\bar{\omega}) = G_0.
\end{align}

The contribution is temperature-independent and does not depend on interaction parameters inside the wire.

\subsection{Dilute soliton gas approximation}

At higher temperatures $T\gg \Delta$ the one-kink approximation is not valid: a larger number of kinks or anti-kinks can be activated. In order to extend the theory to higher temperatures we assume that the soliton gas is dilute and that we may disregard interactions between solitons.

We describe a configuration of the $N$-soliton gas by collective coordinates $\bm{\xi} = \left\{\xi_k\right\}_{k=1}^N$ and labels $\bm{l}= \left\{	l_k\right\}_{k=1}^N$, where $l_k = K,A$ denotes whether the $k$-th soliton is a kink ($K$) or anti-kink ($A$).     

The asymptotic form of the classical solution is given by
\begin{equation}
\phi_{\bm{l},\bm{\xi}}(x) = \sum\limits_{k=1}^N\phi_{l_k}(r,\xi_k),
\end{equation}
where $\check{\phi}_{K(A)}(r,\xi_k)$ is a classical solution for a (anti-) kink located at $r=\xi_k$.

The Green functions are given by
\begin{multline}
i\mathcal{G}_{\alpha\beta}(t,t') = \frac{1}{\mathcal{Z}}\sum\limits_{N=0}^{\infty}e^{-N\beta \Delta }\\ \times \sum\limits_{\bm{l}} \int \left(\prod\limits_{k=1}^N\mathcal{D}\xi_k \exp\left\{iS[\xi_k]\right\}\right)\; 
\phi_{\bm{l},\bm{\xi}(t_\alpha)}(x)\phi_{\bm{l},\bm{\xi}(t'_\beta)}(x') \label{eqn:Green-gas},
\end{multline}	  
where $\alpha,\beta$ are indices in Keldysh space. The partition function is defined as
	\begin{align}
	\mathcal{Z} &=\;\sum\limits_{N=0}^{\infty}e^{-N\beta \Delta } \sum\limits_{\bm{l}} \int \left(\prod\limits_{k=1}^N\mathcal{D}\xi_k \exp\left\{iS[\xi_k]\right\}\right)	.
	\end{align}	
The integration and summation are straightforward and yield a simple expression,
\begin{align}
\mathcal{Z} = \frac{1}{1+e^{-\beta \Delta}}.
\end{align}

\begin{figure}[t]
	\includegraphics[width=\columnwidth]{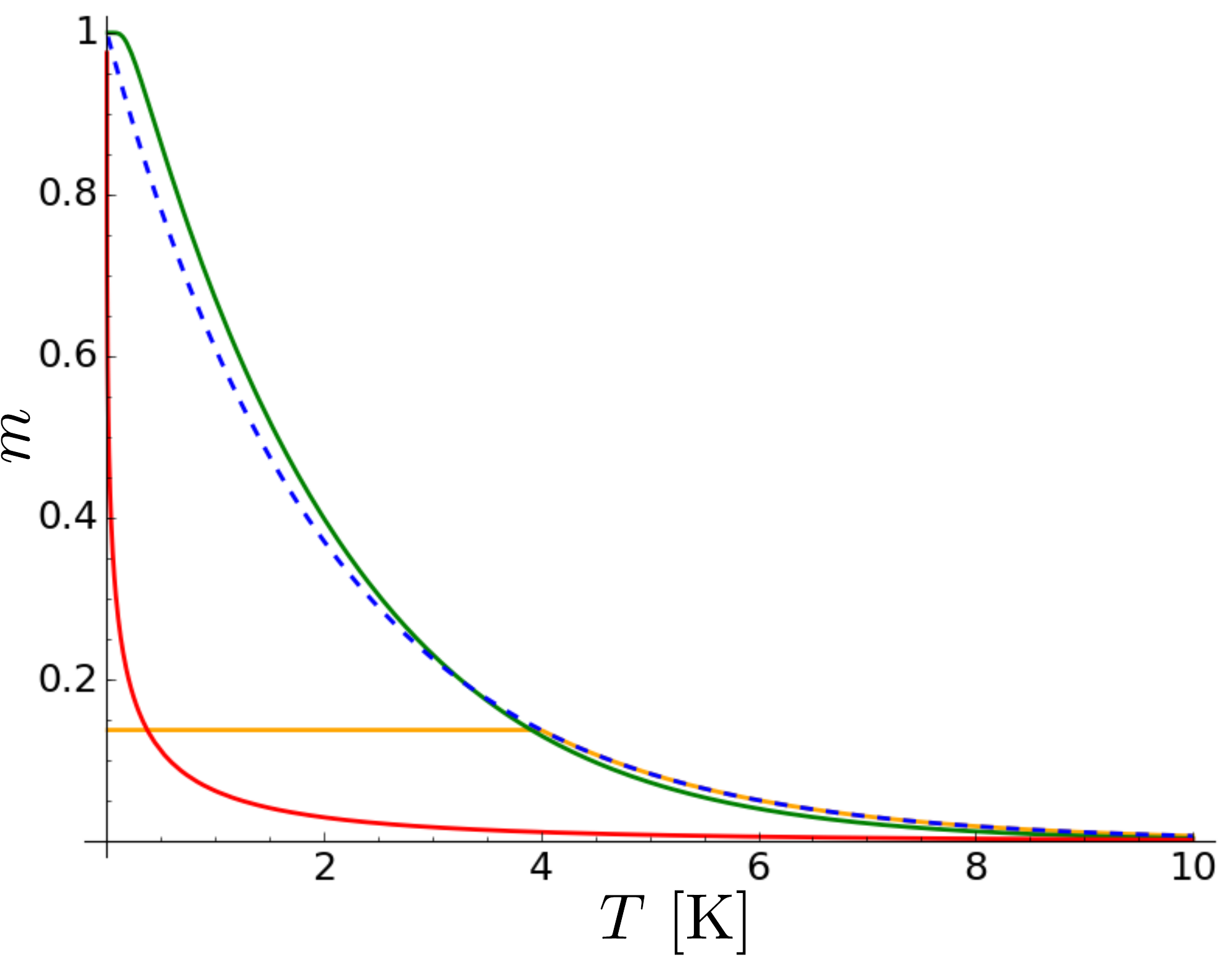}
	\caption{(Color online) Temperature dependence of the order parameter $m(T)$ obtained from: numerical solution of self-consistent condition given by Eq.~(\ref{eqn:order}), see Appendix~\ref{app:order}; exponential approximation by Eq.~(\ref{eqn:order-exp}) with $\nu=1$ (dashed blue line); the same exponential approximation under the assumption that nuclear spin temperature differs from the electron one and is given by~Eq.~(\ref{eqn:nuclear-temperature}) (orange curve); numerical solution of the self-consistent equation~(\ref{eqn:order}) (green line); stretched exponential law~(\ref{eqn:stretched}) with fitting parameters $\nu=0.34$, $T_0=0.05\;\text{T}$. We take the interaction parameter $K_\rho=0.15$, the length of the wire $L=2\;\mu\text{m}$, the hyperfine constant $A=90\;\mu\text{eV}$, number of the nuclear spins in the cross-section of the wire $N_\perp=1000$, and the nuclear spin $I=3/2$.
	}
	\label{fig:order}
\end{figure}

The integrations over $\xi_k$ in~Eq.~(\ref{eqn:Green-gas}) can be reduced to one-kink Green functions $\mathcal{G}_{\alpha \beta}^{(1)}$, and the summation over $N$ can be easily performed,
\begin{align}
i\mathcal{G}_{\alpha\beta}(t,t') = -2\frac{d \ln \mathcal{Z}}{d\left(\beta \Delta\right)}\mathcal{G}^{(1)}_{\alpha \beta} = \frac{2e^{-\beta \Delta}}{1+ e^{-\beta \Delta}}\mathcal{G}_{\alpha \beta}^{(1)}.
\end{align}
In comparison to the one-kink approximation the conductance acquires an extra factor $\left( 1 + e^{-\Delta/T} \right)^{-1}$, and the total conductance is given by
\begin{align}
G= G_0 + \frac{2G_0e^{-\Delta/T}}{1+e^{-\Delta/T}}\sqrt{\frac{T}{\Delta}}\frac{c}{\eta L}e^{-\Delta/T}
\label{eqn:total-conductance-low},
\end{align}
when $\eta \tau_\infty \gg 1$, and the one-soliton expression for crossover to the non-interacting conductance given by~Eq.~(\ref{eqn:one-soliton-result}) is replaced by
\begin{align}
G = G_0 + \frac{2G_0e^{-\Delta/T}}{1+e^{-\Delta/T}} \sqrt{T M} \frac{\dfrac{1-e^{-\eta \tau_\infty}}{M \eta} }{\sqrt{\frac{2}{\pi M} T \tau_\infty^2 \arctan \frac{1}{\eta \tau_\infty}}} 
\label{eqn:total-conductance}.
\end{align}

\section{Temperature dependence of the gap \label{sec:gap}}
We have derived the conductance for a fixed value of the gap $\Delta$. However, the gap itself is temperature dependent and is given by~Eq.~(\ref{eqn:gap}). The Overhauser field also depends on the temperature~\cite{BrauneckerPRB2009,MengEPLJB2014},
\begin{align}
B_{Ov} = \frac{IAm(T)}{2},
\end{align}
where $m(T)$, with $0\le m \le 1$,  is an order parameter. At zero temperature, the nuclear spins are polarized with $m=1$, while at temperatures above $T_c$, $T\gg T_c$, the order is destroyed, and $m = 0$. The order parameter at temperatures near $T_c$ can be estimated as (see Appendix~\ref{app:order} for the derivation)
\begin{align}
m \simeq \exp\left\{ - \frac{1}{\zeta} \left(\frac{T}{T_c} \right)^\nu \right\},
\label{eqn:order-exp}
\end{align}
where the factor $\zeta$ and the exponent $\nu$ are given by~Eqs.~(\ref{eqn:zeta}) and~(\ref{eqn:lambda}). The values of $T_c$, $\nu$, and $\zeta$ for different interaction parameters are shown in  Table~\ref{table:parameters}. The renormalized energy $\varepsilon'_L = \frac{\hbar v'}{L}$ associated with the finite length $L$ turns out to be greater than $T_c$ for every physical value of $K_\rho$, and, hence, the exponent $\nu$ is equal to unity,
\begin{align}
m \simeq \exp\left\{ - \frac{T}{\zeta T_c} \right\}.
\label{eqn:order-exp-unity}
\end{align}
The approximate analytical solution~Eq.~(\ref{eqn:order-exp-unity}) and the numerical solution of the self-consistent condition~Eq.~(\ref{eqn:order}) are shown in  Fig.~\ref{fig:order}.

\begin{table}[b]
	\begin{tabular}{|c|c|c|c|c|c|}
		\hline
		$K_\rho$&g&$T_c$&$\Delta(T=0)$&$\varepsilon'_L$&$\zeta$\\ \hline
		0.10&0.14&0.4 K&0.43 meV, 5.0 K&0.58 meV, 6.7 K&8.8\\ \hline
		0.15&0.21&0.2 K&0.42 meV, 4.8 K&0.38 meV, 4.5 K&10.2\\ \hline
		0.20&0.27&0.1 K&0.40 meV, 4.6 K&0.29 meV, 3.4 K& 10.4\\ \hline
		0.30&0.40&0.02 K&0.36 meV, 4.2 K&0.20 meV, 2.3 K& 9.1\\ \hline
	\end{tabular}	
	\caption{Parameters of the systems calculated for different LL interaction strengths $K_\rho$. The length of the wire is taken to be $L=2\; \text{$\mu$m}$, the hyperfine constant $A=90\;\mu\text{eV}$, the nuclear spin $I=3/2$, the number of spins in the cross-section $N_\perp = 1000$. Critical temperatures are taken from~Ref.~\onlinecite{MengEPLJB2014}. The energy $\varepsilon'_L = \hbar v'/ L$ associated with the finite length $L$ turns out to be greater than $T_c$ for every physically meaningful value of $K_\rho$, and, hence, the exponent $\nu$ in~Eq.~(\ref{eqn:order-exp}) is equal to unity.}	
	\label{table:parameters}
\end{table}

\begin{figure*}
	\includegraphics[width=\linewidth]{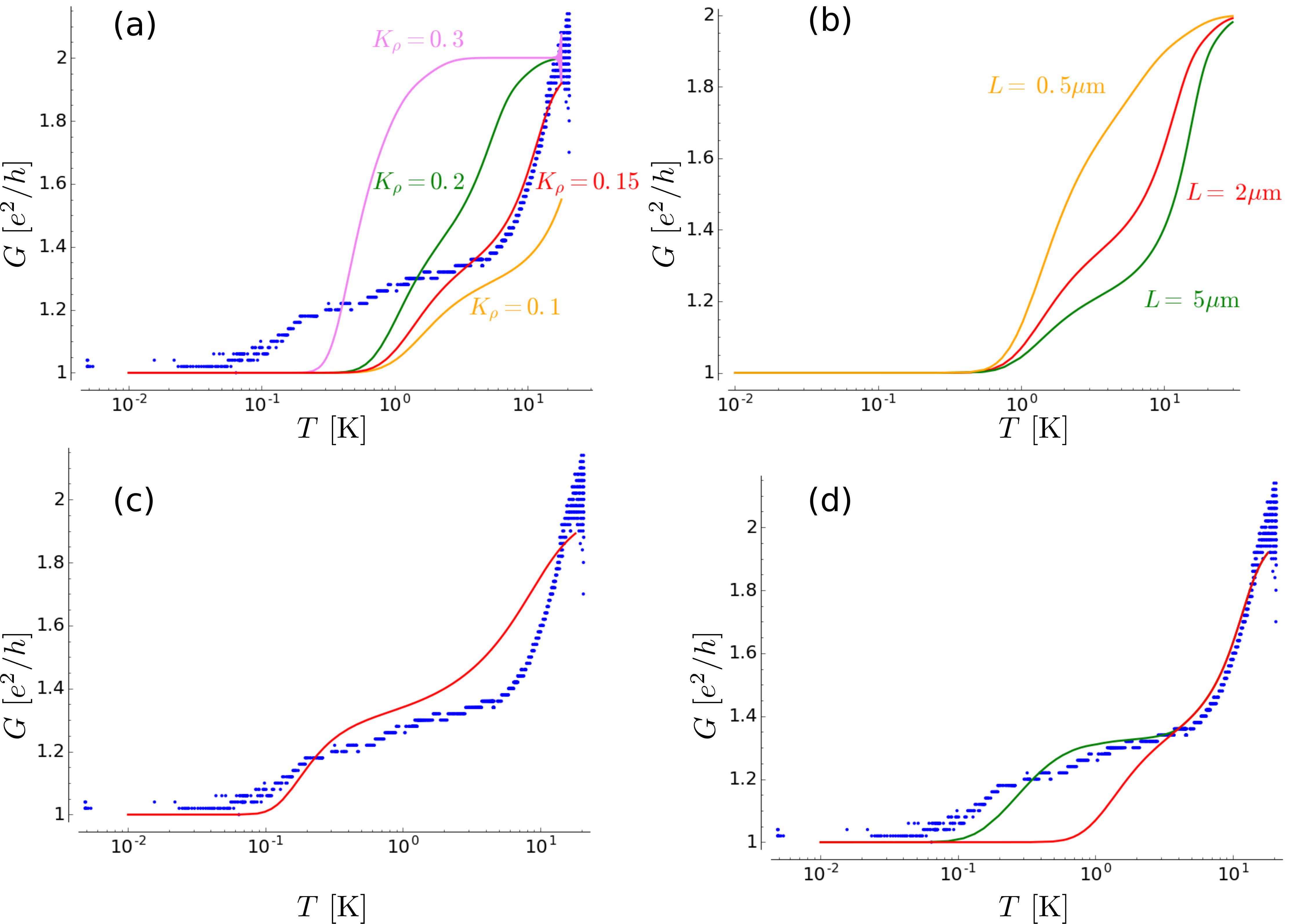}
	\caption{(Color online) Temperature dependence of the conductance $G$ of the wire with helical nuclear spin order. The blue dots show the experimental 
	data~\cite{SchellerPRL2014}.
			(a) $T$-dependence plotted for different $K_\rho$ values, and the order parameter $m$ is assumed to exponentially decay at temperatures above $T_c$ (according to~Eq.~\ref{eqn:order-exp-unity}). We used $L=2\;\mu\text{m}$.
			(b) As in (a) for $K_\rho=0.15$ but for three different values of  $L$.
			(c) $G$ for $K_\rho=0.15$, $L=2\;\mu\text{m}$, $m$ is fitted with the stretched exponential law  Eq.~(\ref{eqn:stretched}) with fit values $T_0\approx 0.05\text{ K}$, $\nu=0.34$.
			(d) $G$ for the case when the nuclear spins are thermally decoupled from the electrons, and $T_N$ is given by~Eq.~(\ref{eqn:nuclear-temperature}) with $T_N^*=4\;\text{K}$ (green curve). The red curve is for $K_\rho=0.15$ and $T_N=T_e$.
			For all plots we used, $A=90\;\mu\text{eV}$, $N_\perp = 1000$, and $I=3/2$.}
	\label{fig:conductance}
\end{figure*}

\section{Discussion and conclusions \label{sec:discussion}}

At low temperatures when the order parameter $m\approx 1$, the gap turns out to be much larger than temperature: for $I=3/2$, $A=90\;\mu\text{eV}$, $K_\rho=0.2$ the temperature associated with the gap $\Delta(T=0)/k_B$ is of order $5\;\text{K}$, while $T_c$ is much lower~(see~Table~\ref{table:parameters}). Under such conditions the conductance determined by~Eq.~(\ref{eqn:total-conductance-low}) is reduced by a factor of $2$ and equals $G_0=e^2/h$. 

At temperatures near $T_c$ the order parameter $m$ and the gap drops, and the conductance features an activation behavior. However, the Ohmic friction discussed in~Sec.~\ref{sec:one-kink} gives rise to a resistivity and causes an additional suppression of the conductance which is now length-dependent~(see~Fig.~\ref{fig:conductance}b). This length prediction could be tested experimentally.
The result resembles the length-dependent suppression of the conductance in the 1D wire with Rashba SOI previously predicted  in~Ref.~\onlinecite{SchmidtPRB2014}. However, in~Ref.~\onlinecite{SchmidtPRB2014} the results were obtained in the limit of weakly interacting electrons, while strong electron-electron interaction is essential for the helical nuclear order considered here.    

 At higher temperatures the order parameter and, hence, the partial gap vanish, and the conductance rises to $2G_0$ as expected. This occurs if $\eta \tau_\infty \ll 1$, {\it i.e.}, if $\Delta(T) T\ll \left(\dfrac{\hbar c}{L} \right)^2$.
  Thus, in the intermediate temperature regime the suppression of the conductance leads to a relatively wide plateau, in qualitative agreement with 
  experiment~\cite{SchellerPRL2014}, see Fig.~\ref{fig:conductance}a.

In Fig.~\ref{fig:conductance}a we plot the conductance obtained with~(\ref{eqn:total-conductance}) for different $K_\rho$, assuming that the order parameter $m(T)$ obeys an exponential law~(\ref{eqn:order-exp-unity}) and compare the result with the experimental data by Scheller~\textit{et al.}\cite{SchellerPRL2014} (blue line in~Fig.~\ref{fig:conductance}a). The theoretical curve shows a good agreement with the experiment at high temperatures for $K_\rho=0.15$. Although at low temperatures the results quantitatively do not agree, the theoretical curve features a conductance plateau at $T\gtrsim T_c$ like the experimental one. 

The quantitative difference can be explained by a suppression of order parameter at low temperatures. Although the theory developed in~Refs.~\onlinecite{BrauneckerPRB2009},~\onlinecite{MengEPLJB2014} allows one to roughly estimate a critical temperature, the exact calculation of order parameter is a subtle issue. 
We conjecture that the order parameter is governed by a more general stretched exponential law similar to~Eq.~(\ref{eqn:order}), but with different $\nu$:
\begin{align}
m = \exp\left\{ -\left( \frac{T}{T_0} \right)^\nu\right\},
\label{eqn:stretched}
\end{align}
and treat $T_0$ and $\nu$ as fitting parameters. The resulting theoretical curve determined by~Eq.~(\ref{eqn:total-conductance}) is close to the experimental one at $\nu\approx 0.34$, $T_0 \approx 0.05\;\text{K}$ (see Fig.~\ref{fig:conductance}c).

Another possible explanation for the seemingly suppressed order at low temperature is as follows. It is conceivable that the electrons and nuclear spins are not in thermodynamic equilibrium, so that the electron and nuclear spin subsystems have different temperatures $T_e$ and $T_N$, respectively, and the measured temperature corresponds to the electron temperature $T_e$, while $T_N$ determines the order parameter $m(T_N)$. As a simple model, we assume that $T_e$ and $T_N$ coincide when the nuclear temperature $T_N$ is higher than some value $T_N^*$, but when the measured (electron) temperature $T_e$ goes below $T_N^*$, the nuclear spin subsystem becomes thermally decoupled from the electrons and stops cooling down, so that the nuclear spin temperature remains at the constant value $T_N^*$. Thus, the order parameter given by Eq.~(\ref{eqn:order-exp-unity}) is determined by the nuclear spin temperature
\begin{align}
T_N(T_e) = T_e\Theta(T_e-T_N^*) + T_N^*\Theta(T_N^*-T_e).
\label{eqn:nuclear-temperature}
\end{align}
The order parameter plotted for different electron temperatures is shown in~Fig.~\ref{fig:order} (orange curve). The temperature dependence of the conductance for this model with $K_\rho=0.15$ and $T_N^*=4\;\mathrm{K}$ is shown in~Fig.~\ref{fig:conductance}d.

For a more reliable comparison between theory and experiment one has to correlate the conductance with the partial gap measured directly at the same temperature, which is yet to be done. 
It is also necessary to point out that in our study we did not take into account the possibility of formation of two or more nuclear spin helices with different directions at temperatures above the critical, which would form domain walls and likely suppress the conductance.

\section*{Acknowledgments}
We would like to thank C.~P.~Scheller, D.~M.~Zumb{\"u}hl, C.-H.~Hsu, P.~Stano, and L.~Glazman for helpful discussions. This work was supported by the Swiss National Science Foundation and by the NCCR QSIT.

\appendix
\section{Fermionic representation\label{app:fermionic}}
From the Hamiltonian density  Eq.~(\ref{eqn:Hamiltonian-lin}) we obtain the equations of motion for the gapped modes ${R}_\up$, ${L}_\down$,
\begin{align}
i\partial_t {R}_\up &= -iv_F \partial_x {R}_\up + b(x) {L}_\down \label{eqn:motion1},\\
i\partial_t {L}_\down &= +iv_F \partial_x {L}_\down + b(x) {R}_\up.\label{eqn:motion2}
\end{align}
In the general case, the Overhauser field depends on position and vanishes in the leads. We search for a solution for electrons incident from the left lead with the following asymptotic forms in the leads,
\begin{align}
&\begin{pmatrix}
R_\up\\ L_\down
\end{pmatrix}e^{i\varepsilon t} = 
\begin{pmatrix}
1\\0 
\end{pmatrix}e^{i\varepsilon \frac{x}{v_F}} + 
s_{LR}
\begin{pmatrix}
0\\ 1
\end{pmatrix}e^{-i\varepsilon\frac{x}{v_F}},\; x\to-\infty,\\
&\begin{pmatrix}
R_\up\\ L_\down
\end{pmatrix} = 
s_{RR}\begin{pmatrix}
1\\0 
\end{pmatrix}e^{i\varepsilon x/v_F}e^{-i\varepsilon t} ,\;x\to+\infty,
\end{align}
where $s_{LR}$ and $s_{RR}$ are scattering amplitudes. The conductance is determined by the transmission coefficient $\mathcal{T} = |s_{RR}|^2$, see Eq.~(\ref{eqn:Landauer}).

\subsection{Abrupt coordinate dependence of Overhauser field \label{ssec:abrupt}}

\begin{figure}
	\includegraphics[width=0.8\columnwidth]{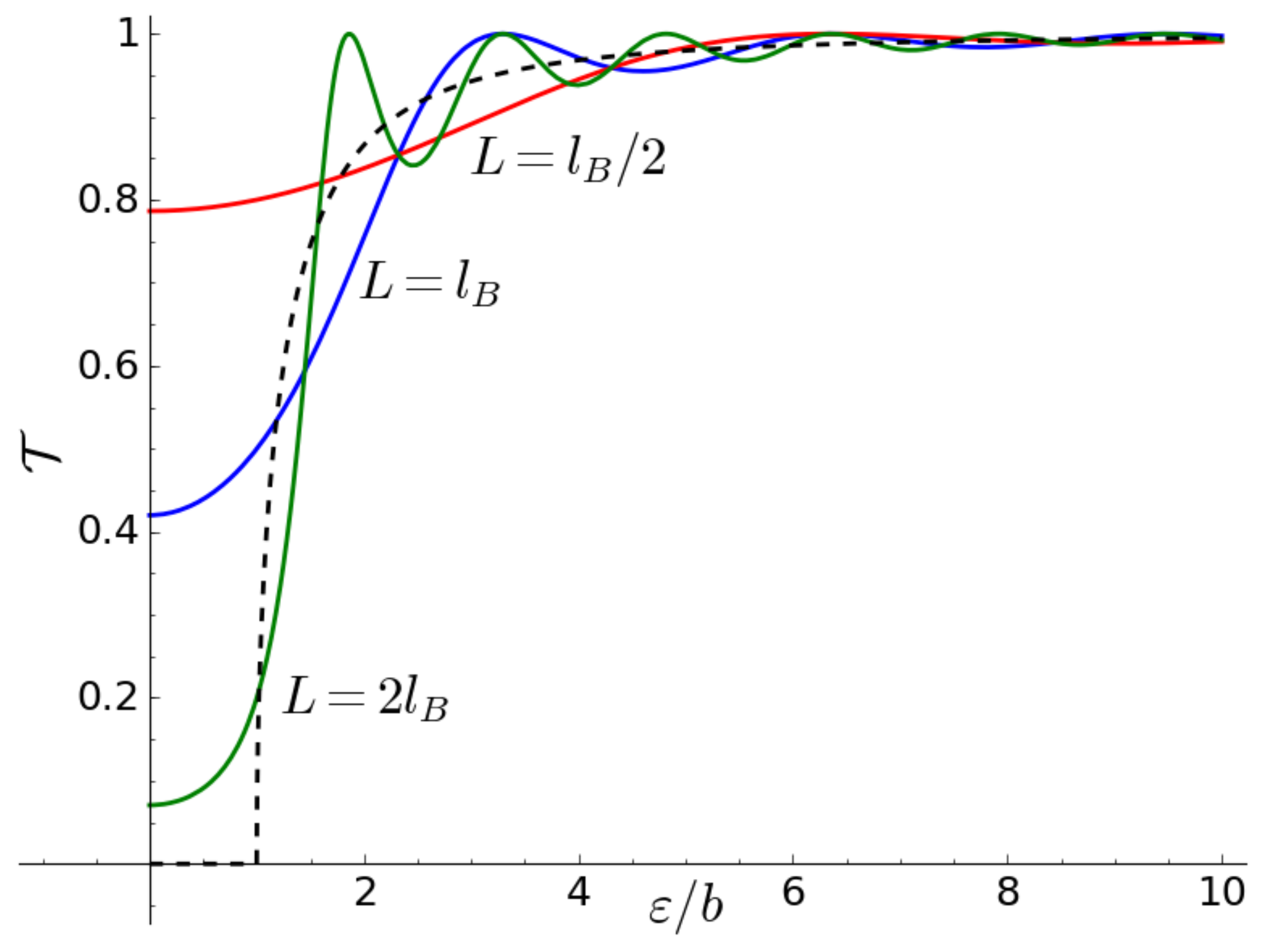}
	\caption{(Color online) Transmission coefficients $\mathcal{T}(\varepsilon)$ given by Eq.~(\ref{eqn:transmission}) for abrupt coordinate dependence of magnetic field on coordinate. The red, green and blue solid lines are plotted for different relations between the wire length $L$ and the magnetic length $l_B = \hbar v_F/b$. The dashed black line shows the averaged value of the transmission coefficient $\bar{\mathcal{T}}(\varepsilon)$ in the limit $L\gg l_B$. }
	\label{fig:transmission-abrupt}
\end{figure}

First, we take $b(x) = b\Theta(x)\Theta(L-x)$.
Solving the equations of motion~(\ref{eqn:motion1})--(\ref{eqn:motion2}), we straightforwardly obtain the transmission coefficient,
\begin{align}
\mathcal{T}(\varepsilon) =
\left\{
\begin{aligned}
 \frac{b^2 - \varepsilon^2}{ b^2 \cosh^2 \left(q L\right) - \varepsilon^2},\, |\varepsilon|<b,\\
 \frac{\varepsilon^2 - b^2}{\varepsilon^2 - b^2 \cos^2\left(q L\right)},\,|\varepsilon|>b,
\end{aligned}
 \right. 
 \label{eqn:transmission}
\end{align}
where $q=\sqrt{|\varepsilon^2-b^2|}/v_F$. The transmission coefficient depends on the relation between the wire length $L$ and the length associated with the Overhauser field $l_B = \hbar v_F/b$ (see Fig.~\ref{fig:transmission-abrupt}).

\begin{figure}[b]
	\includegraphics[width=0.8\columnwidth]{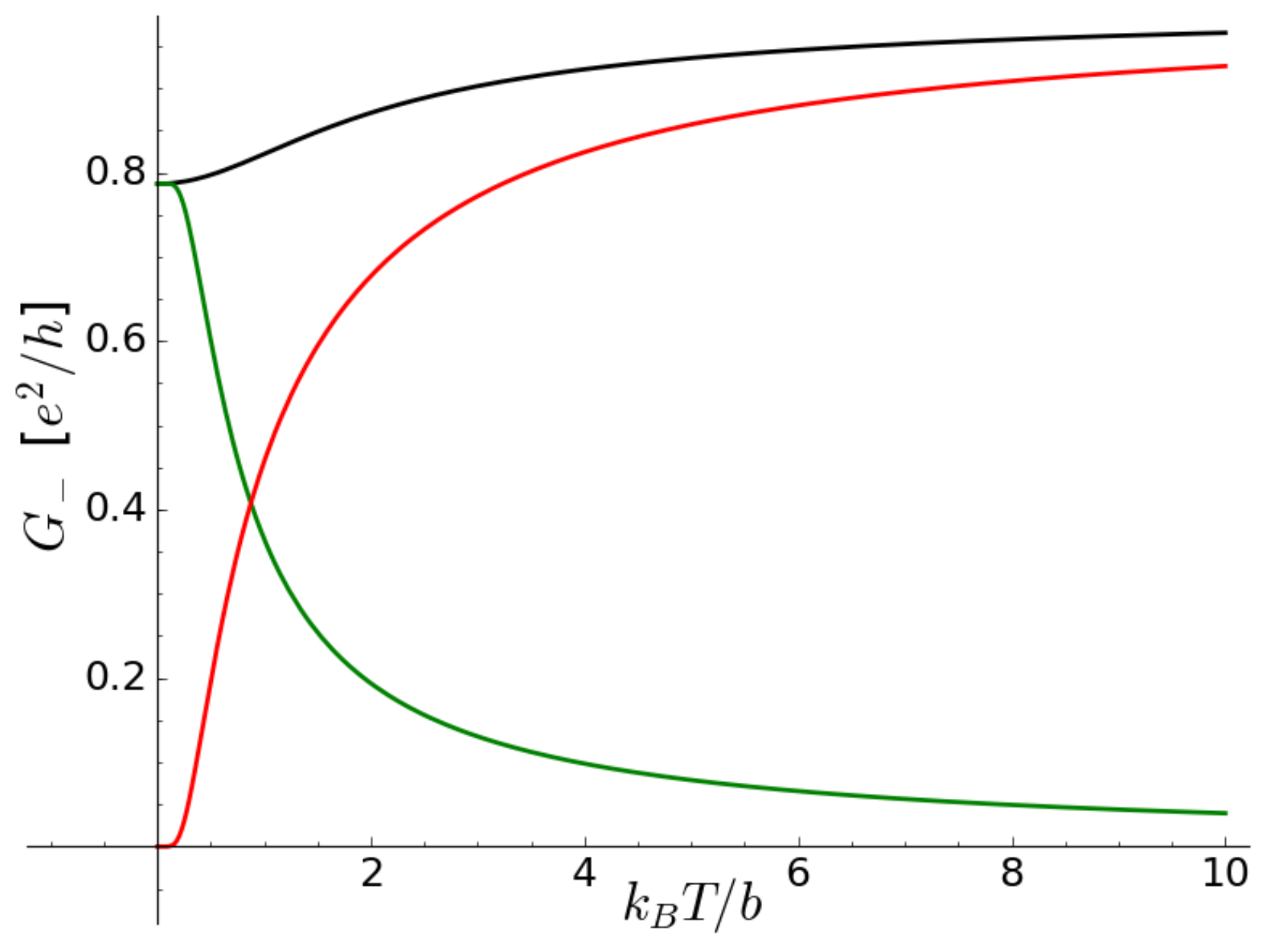}
	\caption{(Color online) Contribution to the conductance from the gapped mode (black line) in non-interacting approximation consisting of tunneling contribution~(green line), contribution from the states above the gap~(red line). The length of the wire is taken to be comparable to the magnetic length $l_B$, $L=l_B/2$. At temperatures much higher than $b$ the tunneling contribution becomes negligible.}
	\label{fig:tunneling}
\end{figure}

The conductance $G$ can now be calculated by using the generalized Landauer formula, see~Eq.~(\ref{eqn:Landauer}). If $L>l_B$, the tunneling and Fabry-Perot oscillations can be disregarded and the transmission coefficient can be replaced with its averaged value $\bar{\mathcal{T}}(\varepsilon) = \sqrt{\varepsilon^2-b^2}/\varepsilon$ (see Fig.~\ref{fig:transmission-abrupt}). If $L$ is comparable with $l_B$, then the tunneling becomes significant at low temperatures $T< b$. At higher temperatures the main contribution is again from thermally excited electrons above the gap (see Fig.~\ref{fig:tunneling}).

At temperatures below the critical one, $T<T_c$, for typical values of hyperfine constant $A=90\;\mu\text{eV}$, Fermi velocity $v_F=2\cdot10^5\;\mathrm{m/s}$, short-length cutoff $a= 5.65\;\mathrm{\AA}$, and the interaction parameter $K_\rho\sim 0.2$,  the half-gap can be estimated as $b=0.4 \text{meV}$ (see Table~\ref{table:parameters}), and the associated magnetic length $l_B \sim 0.4\;\mathrm{\mu m}$. Thus, if the wire length $L\gtrsim 1\;\mathrm{\mu m}$, the tunneling contribution at low temperatures is negligible. At higher temperatures $T>b(T)$, when the conductance manifests activation-law behavior, the contribution from the tunneling can also be disregarded.

\subsection{Smooth coordinate dependence of the Overhauser field $B(x)$ \label{ssec:smooth}}

Now we assume that the Overhauser field $B(x)$ depends on position $x$, varying from zero in the leads to some finite value $B$ inside the wire. If $B(x)$ varies slowly over the distances comparable with $l_{B}$, the reflections at the contacts (between leads and wire) can be neglected and the transmission coefficient becomes $\mathcal{T}=\Theta(|\varepsilon|-b)$. The generalized Landauer formula, see Eq.~(\ref{eqn:Landauer}), yields 
\begin{align}
G_-=G_0\left(1-\tanh \frac{b}{2T}\right).
\end{align}

\begin{figure}
	\includegraphics[width= 0.8 \columnwidth]{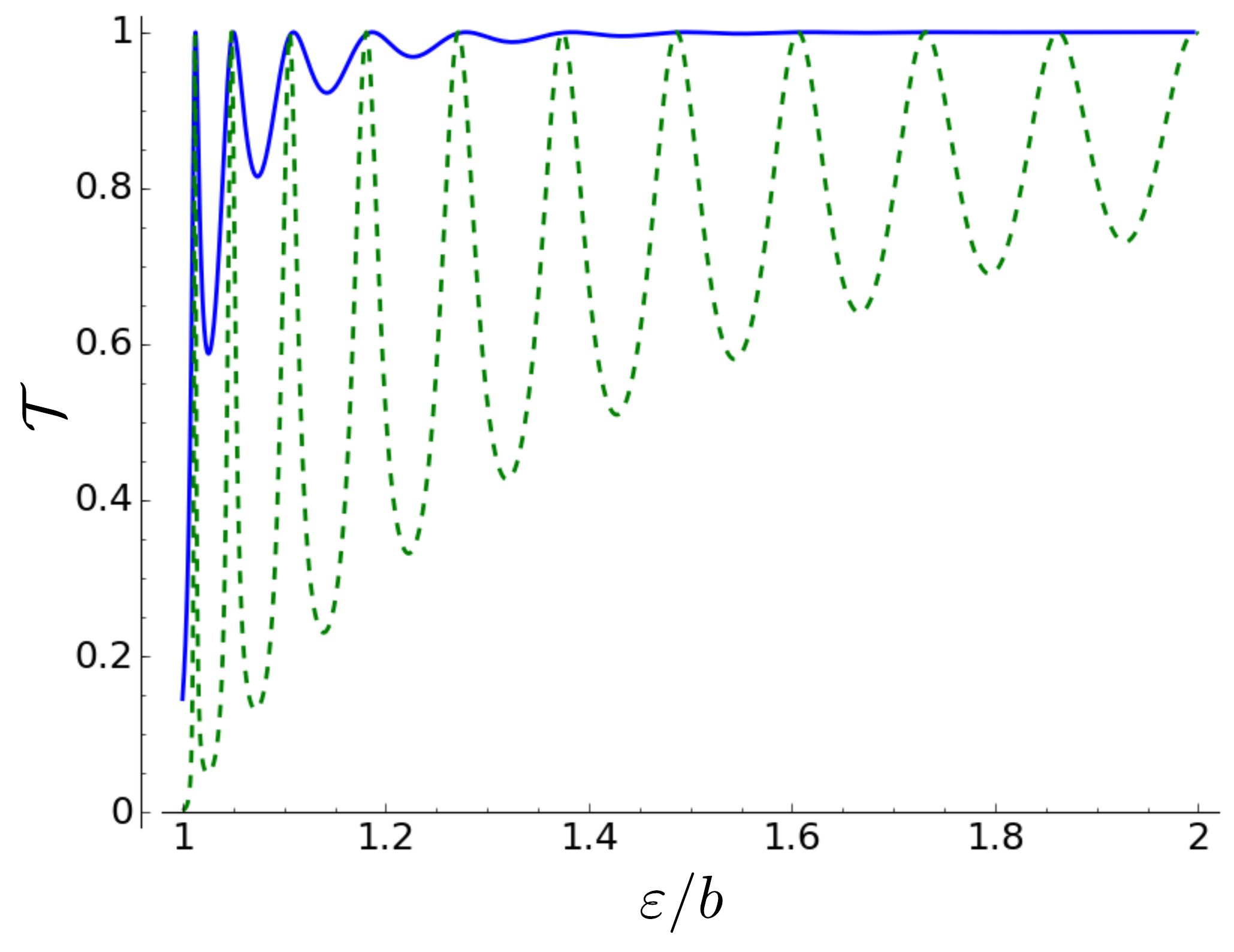}
	\caption{(Color online) Transmission coefficients $\mathcal{T}(\varepsilon)$ in case of coordinate-dependent Overhauser field $B(x)$ for a long wire $L=20 l_{B}$. Blue solid line: the coordinate dependence of Overhauser field $B(x)$ is given by~Eq.~(\ref{eqn:B-tanh}) with $l=l_B$. Green dashed line: the Overhauser field drops abruptly at the contacts, $l\to 0 $.}
	\label{fig:transmission-profile}
\end{figure}

In order to study the more general case we solve the equations of motion~(\ref{eqn:motion1})--(\ref{eqn:motion2}) numerically by using the 4th order Runge-Kutta method and by assuming that $B(x)$ is given by
\begin{equation}
B(x) = \frac{B}{2}\left(\tanh \frac{x}{l} + \tanh \frac{L-x}{l} \right).
\label{eqn:B-tanh}
\end{equation}
The resulting transmission coefficients are shown in~Fig.~\ref{fig:transmission-profile} (blue line). The resulting conductance $G$ calculated using Eq.~(\ref{eqn:Landauer}) is plotted in~Fig.~\ref{fig:nonint-conductance} (blue line).

\subsection{Parabolic electron dispersion}

\begin{figure*}
	\includegraphics[width=\linewidth]{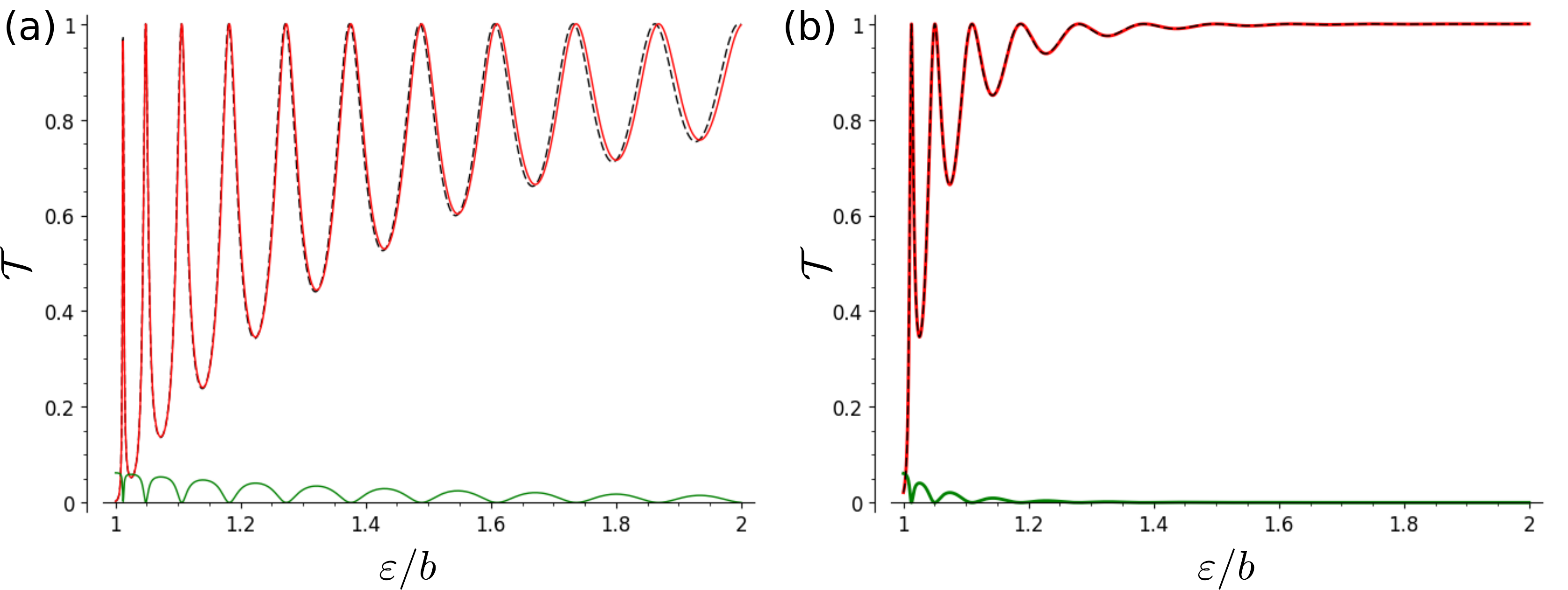}
	\caption{(Color online) Transmission coefficients $\mathcal{T}_{\up\up}$ (red line) and $\mathcal{T}_{\down\up}$ (green line) for a long wire, $L=20l_b$ with parabolic electron dispersion~(with $\varepsilon_F = 100b$). The dashed black line corresponds to the transmission coefficient $\mathcal{T}$ obtained in Secs.~\ref{ssec:abrupt}--\ref{ssec:smooth} for a linearized dispersion. (a) The Overhauser field drops abruptly at the contacts. (b) The coordinate dependence of Overhauser field $B(x)$ is given by Eq.~(\ref{eqn:B-tanh}) with $l=l_B$.}
	\label{fig:parabolic-transmissions}
\end{figure*}

\begin{figure*}
	\includegraphics[width=\linewidth]{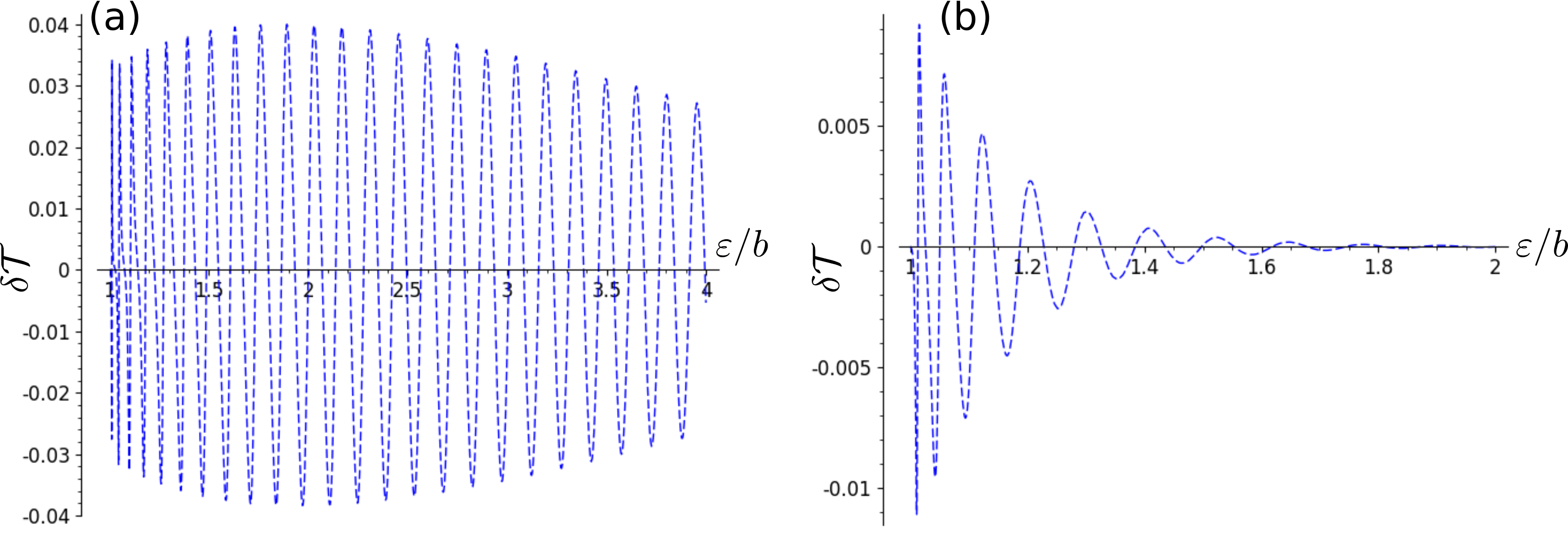}
	\caption{(Color online) Difference $\delta \mathcal{T} = \mathcal{T}_{\up\up} -\mathcal{T}$ between the transmission coefficient $\mathcal{T}_{\up\up}$ for parabolic electron dispersion~(with $\varepsilon_F = 100b$) and the transmission coefficient $\mathcal{T}$ obtained in Secs.~\ref{ssec:abrupt}--\ref{ssec:smooth}. (a) The Overhauser field drops abruptly at the contacts. (b) The coordinate dependence of Overhauser field $B(x)$ is given by Eq.~(\ref{eqn:B-tanh}) with $l=l_B$.}
	\label{fig:transmissions-difference}
\end{figure*}

Previously we used a Hamiltonian given by Eq.~(\ref{eqn:Hamiltonian-lin}) with a spectrum of electrons linearized near the Fermi points. In this section we calculate numerically the conductance of a non-interacting wire with a helical Overhauser field for a parabolic electron dispersion. 

We start from the 
eigenvalue equation
in energy representation obtained from the Hamiltonian given by~Eq.~(\ref{eqn:Hamiltonian}),
\begin{align}
\varepsilon \psi_s = -\frac{\partial^2_x -k_F^2}{2m}\psi_s + b(x) \psi_{-s}e^{is2k_F x}.
\label{eqn:motion-quadratic}
\end{align}
It is convenient to introduce new variables $\tilde{\psi}_s = \psi_s e^{-isk_F x}$ to gauge away the fast oscillating terms in  Eq.~(\ref{eqn:motion-quadratic}). Then the equations for the new variables read,
\begin{align}
\varepsilon \tilde{\psi}_s = -\frac{\partial^2_x}{2m}\tilde{\psi}_s -isv_F \partial_x\tilde{\psi}_s + b(x)\tilde{\psi}_{-s}.
\end{align}
Although these eigenvalue equations resemble Eqs.~(\ref{eqn:motion1})--(\ref{eqn:motion2}) which were obtained for the gapped mode, their solutions describe now both gapped and gapless modes.

In order to find the scattering amplitudes for waves incident from the left lead, $x<0$, we have to impose  proper boundary conditions. The  wavefunctions of spin-up electrons moving from the left to the right satisfying the eigenvalue equation are of the form,
\begin{align}
&\psi_\up(x<0,\varepsilon) = e^{ikx} + r_{\up \up} e^{-ikx},\quad \psi_\down(x<0,\varepsilon) = r_{\down\up}e^{-ikx}, \label{eqn:solution-left-lead}\\
&\psi_\up(x>L,\varepsilon) = t_{\up\up}e^{ik(x-L)}, \quad \psi_\down(x>L,\varepsilon) = t_{\down\up}e^{ik(x-L)},
\label{eqn:solution-right-lead}
\end{align}
where $k=\sqrt{2m(\varepsilon+\varepsilon_F)}$, $t_{s's}$ and $r_{s's}$ are transmission and reflection amplitudes. If the wavefunction has the general form $\psi = 
\alpha_{+}e^{ikx} + \alpha_{-} e^{-ikx}$,
then the amplitudes for right- and left-movers, $\alpha_{+}$, $\alpha_{-}$, can be expressed in terms of $\psi$ and its spatial derivative $\psi'$ at the boundary,
\begin{align}
\alpha_{+} = \frac{ik\psi(x=0^-) + \psi'(x=0^-)}{2ik},\\
\alpha_{-} = \frac{ik\psi(x=0^-) - \psi'(x=0^-)}{2ik}.
\end{align}

Then the solution in the left lead, given by Eq.~(\ref{eqn:solution-left-lead}), obeys the following boundary conditions,
\begin{align}
 \frac{ ik\psi_\up(x=0^-) + \psi'_\up(x=0^-)}{2ik} = 1,\\
ik\psi_\down(x=0^-) +  \psi'_\down(x=0^-) = 0,
\end{align}
and the boundary conditions for the solution in the right lead, given by Eq.~(\ref{eqn:solution-right-lead}), are
\begin{align}
 ik \psi_\up(x=L^+) - \psi'_\up(x=L^+) = 0,\\
ik \psi_\down(x=L^+) -  \psi'_\down(x=L^+) = 0,
\end{align}
where $L^{\pm}=L\pm 0$.

The continuity conditions for the wavefunctions $\psi_s$ and their spatial derivatives $\psi'_s$ yield the following relation between  $\psi$ and $\tilde{\psi}$,
\begin{align}
&\psi_s(x=0^-) = \tilde{\psi}_s(x=0^+)\\
&\psi'_s(x=0^-) = \tilde{\psi}'_s(x=0^+) +ik_Fs \tilde{\psi}_s(x=0^+)\\
&\psi_s(x=L^+)e^{-ik_FLs} = \tilde{\psi}_s(x=L^-)\\
&\psi'_s(x=L^+)e^{-ik_FLs} = \tilde{\psi}'_s(x=L^-) +ik_Fs \tilde{\psi}_s(x=L^-).
\end{align}

Finally, the boundary conditions for the wire read,
\begin{align}
&i(k+k_F)\tilde{\psi}_{\up}(x=0^+) + \tilde{\psi}_\up'(x=0^+) = 2ik,\label{eqn:quadratic-bc1}\\
&i(k-k_F)\tilde{\psi}_{\down}(x=0^+) + \tilde{\psi}_\down'(x=0^+) = 0,\\
&i(k-k_F)\tilde{\psi}_{\up}(x=L^-) - \tilde{\psi}_\up'(x=L^-) = 0,\\
&i(k+k_F)\tilde{\psi}_{\down}(x=L^-) - \tilde{\psi}_\down'(x=L^-) = 0\label{eqn:quadratic-bc4}.
\end{align}
Similarly, the boundary conditions for the wavefunction of incident spin-down  electrons read,
\begin{align}
&i(k+k_F)\tilde{\psi}_{\up}(x=0^+) + \tilde{\psi}_\up'(x=0^+) = 0,\label{eqn:quadratic-bc5}\\
&i(k-k_F)\tilde{\psi}_{\down}(x=0^+) + \tilde{\psi}_\down'(x=0^+) = 2ik,\\
&i(k-k_F)\tilde{\psi}_{\up}(x=L^-) - \tilde{\psi}_\up'(x=L^-) = 0,\\
&i(k+k_F)\tilde{\psi}_{\down}(x=L^-) - \tilde{\psi}_\down'(x=L^-) = 0.\label{eqn:quadratic-bc8}
\end{align}

\begin{figure}
	\includegraphics[width=\columnwidth]{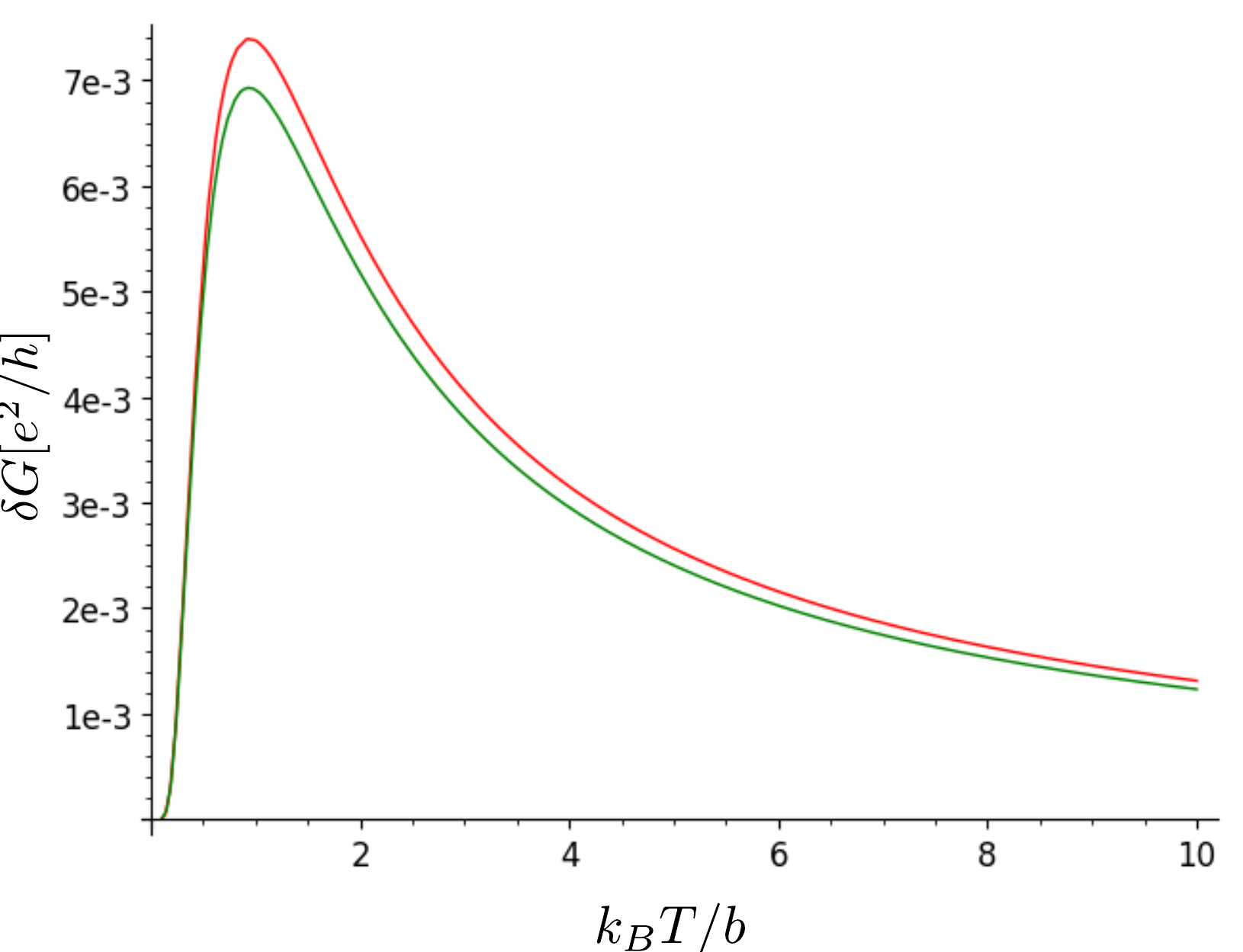}
	\caption{(Color online) Correction $\delta G = G^{par} - G^{lin}$ to the temperature dependence of the conductance due to the nonlinearity of the electron dispersion in case for an abrupt coordinate dependence of the Overhauser field $B(x)$ (red line) and a smooth one
	given by Eq.~(\ref{eqn:B-tanh}) with $l=l_B$ (green line). The main correction is caused by the emergence of a non-zero transmission $\mathcal{T}_{\down\up}$ for the quadratic dispersion, which vanishes for linear dispersion.}
	\label{fig:conductance-difference}
\end{figure}

We solve~Eq.~(\ref{eqn:motion-quadratic}) with the boundary conditions given by Eqs.~(\ref{eqn:quadratic-bc1})--(\ref{eqn:quadratic-bc4}) and Eqs.~(\ref{eqn:quadratic-bc5})--(\ref{eqn:quadratic-bc8}) numerically, using the fourth order Runge-Kutta method and with $B(x)$ given by~Eq.~(\ref{eqn:B-tanh}). The transmission coefficients $\mathcal{T}_{\up\up}$, $\mathcal{T}_{\down\up}$, defined as  $\mathcal{T}_{ss'} = |t_{s's}|^2$, are plotted in Fig.~\ref{fig:parabolic-transmissions}. In the case of parabolic electron dispersion the gapped and gapless modes are not decoupled, and now there appears a non-zero probability $\mathcal{T}_{\down\up} \sim b/\varepsilon_F$ for a spin-up right-moving electron in the left lead to scatter into a spin-down right-moving electron in the right lead. For the two remaining transmission coefficients we obtain $\mathcal{T}_{\down\down} = 1$, $\mathcal{T}_{\up\down} = 0$ within the precision of our numerics.

From the numerical results~(see Fig.~\ref{fig:transmissions-difference}) we can conclude that the transmission coefficient $\mathcal{T}$ obtained in sections~\ref{ssec:abrupt}--\ref{ssec:smooth} for the linearized dispersion is indeed a good approximation for $\mathcal{T}_{\up\up}$.

The conductance can now be calculated using the generalized Landauer formula similar to~Eq.~(\ref{eqn:Landauer}),
\begin{align}
G = G_0\int\limits_{-\infty}^{+\infty} \frac{\sum_{ss'}\mathcal{T}_{ss'}}{4T\cosh^2(\varepsilon/2T)}.
\end{align}

The difference between the conductances calculated for parabolic (with $\varepsilon_F = 100$) and linear dispersions is shown in Fig.~\ref{fig:conductance-difference}. Thus, we can conclude that the model with linearized electron dispersion yields the correct result for the conductance provided $b,T \ll \varepsilon_F$.

\section{Green function definitions \label{app:Green-def}}
We use the following definitions for the Green functions of bosonic variables (fields $\phi_{\pm}$ and collective coordinates~$\xi$),
\begin{align}
&\mathcal{G}^R_\pm(x,x',t,t') = -i\Theta(t-t')\left\langle[\phi_\pm(x,t),\phi_\pm(x',t')]\right\rangle,\\
&\mathcal{G}^A_\pm(x,x',t,t') = i\Theta(t'-t)\left\langle\left[\phi_\pm(x,t),\phi_\pm(x',t')\right]\right\rangle,\\
&\mathcal{G}^K_\pm(x,x',t,t') = -i\left\langle\left\{\phi_\pm(x,t),\phi_\pm(x',t')\right\}\right\rangle,\\
&\mathcal{G}^M_\pm(x,x',\tau) = -\left\langle \phi_\pm(x,\tau) \phi_\pm(x',0)\right\rangle,\;0<\tau<\beta\\
&\mathcal{D}^R_\pm(t,t') = -i\Theta(t-t')\left\langle[\xi(t),\xi(t')]\right\rangle,\\
&\mathcal{D}^A_\pm(t,t') = i\Theta(t'-t)\left\langle\left[\xi(t),\xi(t')\right]\right\rangle,\\
&\mathcal{D}^K_\pm(t,t') = -i\left\langle\left\{\xi(t),\xi(t')\right\}\right\rangle,\\
&\mathcal{D}^M_\pm(\tau) = -\left\langle \xi(\tau) \xi(0)\right\rangle,\;0<\tau<\beta,
\end{align}
where $\Theta$ denotes the Heaviside step-function, the plus~(minus) index denotes gapless~(gapped) modes.

\section{Effective action for the collective coordinate \label{app:eff-action}}
We follow~Ref.~\onlinecite{BraunPRB1996} in order to derive an effective action for a kink-particle. We work only to order $O\left( (\partial_\tau\xi/c) ^2\right)$ and use the notation $\check{\phi} = \begin{pmatrix}\phi_- ,& \phi_+ \end{pmatrix}^T$. The action can be expanded around the one-kink solution,
$\check{\varphi}(x,\tau) = \check{\phi}^{k}(x-\xi) + \delta\check{\varphi}(x-\xi,\tau)$, where $\delta\check{\varphi}(x,\tau)$ describes fluctuations around the  static classical  path,
	\begin{align}
	&S= \beta M c^2 +S_\xi + S_\varphi,\\ 
	&S_\xi = \frac{1}{2\pi v_F}\int d\tau\; \left[\int dx\; (\partial_x\check{\phi}^{k},{\partial_x\check{\phi}^{k}})\right](\partial_\tau
	\xi)^2,\\
	&S_\varphi =  \frac{1}{2\pi v_F}\int dx d\tau\; \left(\delta\check{{\varphi}},({\mathscr{H}}_0 + {\mathscr{H}_1})  \delta\check{\varphi}\right) + \left(\mathscr{J}, \delta\check{{\varphi}} \right),
	\end{align}
	where we introduced the scalar product, $\left(\mathscr{J}, \delta\check{{\varphi}} \right)={\mathscr{J}}_+  \delta{{\varphi}_+} + {\mathscr{J}}_-  \delta{{\varphi}_-}$, etc.
	The operators $\mathscr{H}_0$, $\mathscr{H}_1$, and the current $\mathscr{J}$ are defined as
	\begin{align}
	&{\mathscr{H}_0} =\partial_\tau^2 +\partial_x\begin{pmatrix}
	v_+^2&v_-^2\\
	v_-^2&v_+^2
	\end{pmatrix}\partial_x - \begin{pmatrix}
	V^2(x)&0\\
	0&0
	\end{pmatrix},\\
	& {\mathscr{H}}_1 =   2(\partial_\tau \xi) \partial_x\partial_\tau - (\partial_\tau \xi)^2\partial_x^2,\\
	&\mathscr{J} = -2\left(\partial_\tau \xi\right)^2 \partial_x^2\check{\phi}^k,
	\end{align}
	and the potential is given by
	\begin{align}
	V^2(x) = \left(\frac{c}{\delta_0}\right)^2\left(1 - 2\mathrm{sech}^2\left(x/\delta_0\right) \right).
	\end{align}  
	
	The effective action for $\xi$ can be represented as
	\begin{align}
	S_{eff}[\xi] = S_\xi -\ln\left\{\int^\prime \mathcal{D}\delta\varphi\; \det \left(\frac{\delta Q}{\delta \xi} \right)e^{-S_\varphi} \right\}.
	\label{eqn:eff-action}
	\end{align}  
	The prime denotes that the the integration is performed over fluctuations orthogonal to the zero-mode $\partial_x\check{\phi}^k$.
In order to integrate out fluctuations we shift $\check{\varphi}$ by $\check{\rho} \equiv (1/2)\mathscr{H}^{-1}\mathscr{J}$, replacing $\check{\varphi} \to \check{\varphi} - \check{\rho}$. 

Similar to~Ref.~\onlinecite{BraunPRB1996}, the Faddeev-Popov (Jacobian) determinant $\det (\delta Q/\delta \xi)$ leads to an extra term in the action proportional to $(\partial_\tau \xi)^2$, which is a (small) mass renormalization and its exact value is not of interest here since the value of the renormalized mass $M=\Delta/c^2$ is given by~Eq.~(\ref{eqn:gap}).

We now turn to the integration over $\delta\check{\varphi}$ in~Eq.~(\ref{eqn:eff-action}),
\begin{align}
\int^\prime \mathcal{D}\delta\varphi\; e^{-S_\varphi} = \frac{1}{\sqrt{\det'\left(\mathscr{H}_0+\mathscr{H}_1 \right)}}.
\end{align}
The prime on the determinant denotes omission of the zero mode. Using the identity $\ln \det = \mathrm{tr} \ln$ we expand
\begin{multline}
 \frac{1}{\sqrt{\det'\left(\mathscr{H}_0+\mathscr{H}_1 \right)}} = \exp\left\{ -\frac{\mathrm{tr}' \ln\left(\mathscr{H}_0\left[ 1+\mathscr{H}_0^{-1}\mathscr{H}_1\right] \right)}{2} \right\}\\
 \approx \frac{1}{\sqrt{\det' \mathscr{H}_0}} \exp\left\{ -\frac{\mathrm{tr}' \left[ \mathscr{H}_0^{-1}\mathscr{H}_1 -\frac{1}{2}\left(\mathscr{H}_0^{-1}\mathscr{H}_1\right)^2\right] }{2} \right\}.
\end{multline}
Since $\mathscr{H}_1 = O\left( \partial_\tau \xi/c\right)$ this represents an expansion in increasing powers of $\partial_\tau X/c$.

Similar to~Ref.~\onlinecite{BraunPRB1996}, the first order term $\mathscr{H}_0^{-1}\mathscr{H}_1$ leads to terms proportional to $\partial_\tau \xi^2$, renormalizing the mass. The second order term $\left(\mathscr{H}_0^{-1}\mathscr{H}_1\right)^2$ is more interesting for our purpose.
The operator ${\mathscr{H}_0}$ describes free mesons. The spectrum of mesons can be found by solving Schroedinger equation for the eigenfunctions at $|x| \gg \delta$ where $\mathrm{sech}^2\left( x/\delta \right)$ vanishes,
\begin{align}
\begin{pmatrix}
v_+^2q^2 + (c/\delta)^2 - \omega_q^2 &v_-^2q^2\\
v_-^2q^2&v_+^2q^2-\omega_q^2
\end{pmatrix}\delta\check{\varphi}_q = 0  .
\end{align}
The fluctuation spectrum consists of two branches, one of which has a gap $c/\delta$ and the other is gapless,
\begin{align}
\left(\omega_q^{\pm}\right)^2 = \frac{(c/\delta)^2+2 q^2 v_+^2\pm\sqrt{(c/\delta)^4+4 q^4 v_-^4}}{2}.
\end{align}

The eigenfunctions of $\mathscr{H}_0$ factorize into a space and time part $\left|\pm,q,\bar{\omega}\right\rangle = \left|q\right\rangle \left|\bar{\omega}\right\rangle$, where $\left\langle \tau \middle| \bar{\omega} \right\rangle = e^{i\bar{\omega} \tau}/\sqrt{\beta}$. 
Using these notations, we have up to order $(\partial_\tau \xi/c)^2$,
\begin{multline}
\frac{1}{4}\mathrm{tr'}\left(\mathscr{H}_0^{-1}\mathscr{H}\right)^2 \\= \sum\limits_{\nu=\pm,q,q',\omega,\omega'}\frac{ \left|\left\langle \nu, q',\bar{\omega}'\middle|(\partial_\tau \xi)\partial_x |\nu,q,\bar{\omega}\right\rangle \right|^2}{\left(\bar{\omega}^2+\left(\omega_q^{\nu}\right)^2\right)\left(\bar{\omega}^2+\left(\omega_{q'}^{\nu}\right)^2\right)}.
\label{eqn:trace-2nd-order}
\end{multline}

In leading order in wire length $L$ we get
\begin{multline}
\left\langle\nu,q,\bar{\omega}\middle|(\partial_\tau \xi)\partial_\tau\partial_x \middle|\nu,q',\bar{\omega}'  \right\rangle =\\ -\frac{q\bar{\omega}'}{\beta}\delta_{qq'}\int d\tau
e^{i(\bar{\omega}'-\bar{\omega})\tau}\partial_\tau \xi(\tau).
\end{multline}

Thus Eq.~(\ref{eqn:trace-2nd-order}) can be rewritten in the form
\begin{align}
\frac{1}{4}\mathrm{tr'}\left(\mathscr{H}_0^{-1}\mathscr{H}\right)^2 = 
T\sum\limits_{\bar{\omega}} \bar{\omega}^2 \xi_{\bar{\omega}} \xi_{-\bar{\omega}} \Gamma(\bar{\omega}),
\end{align}
with the damping kernel
\begin{multline}
\Gamma^\nu(\bar{\omega}) = T\sum\limits_{\bar{\omega}', q}\frac{q^2\left(\bar{\omega}+\bar{\omega}'\right)\bar{\omega}'}{\left[\left(\bar{\omega}+\bar{\omega}'\right)^2+\omega^{\nu2}_q\right]\left[\bar{\omega}^{\prime2}+\omega^{\nu2}_{q}\right]}.
\end{multline}
Performing the summation over Matsubara frequencies $\bar{\omega}_n = 2\pi n T$, we obtain
\begin{align}
\Gamma^\nu(\bar{\omega}) = \sum\limits_q \frac{4q^2\omega^\nu_q \coth\left(\beta \omega^\nu_q/2\right)}{4\left(\omega_q^{\nu}\right)^2+\bar{\omega}^2}.
\end{align}

To render the results finite in the thermodynamic limit, we have to subtract the vacuum fluctuations~\cite{Sakita1985}. This renormalization simply amounts to the replacement~(see~Ref.~\onlinecite{BraunPRB1996} for detailed explanations)
\begin{align}
\sum_{\nu,q}\to \sum_{\nu}\int dq \left[\rho^\nu(q) - \frac{L}{2\pi} \right]=\sum\limits_\nu\int \frac{dq}{2\pi} \frac{2\delta^{-1} v_\nu^2 }{\left(\omega_q^{\nu}\right)^2},
\end{align}
where $\rho^\nu(q)$ is the density of states for the gapped ($\nu=+$) and gapless ($\nu=-$) modes, respectively.

Finally, the damping kernel $\Gamma^\nu$ is given by
\begin{align}
\Gamma^\nu(\bar{\omega}) = \int \frac{dq}{2\pi} \frac{2\delta^{-1} v_\nu^2 }{\left(\omega_q^{\nu}\right)^2}\frac{4q^2\omega^\nu_q \coth\left(\beta \omega^\nu_q/2\right)}{4\left(\omega_q^{\nu}\right)^2+\bar{\omega}^2}.
\end{align}

For the gapped mode, the integration for $\Gamma^+$ does not diverge in the infrared limit, and $\Gamma^+$ is of order $O(\bar{\omega}^0)$. Therefore, the gapped modes contribute only to the mass renormalization.

In order to estimate $\Gamma^-$, we linearize the spectrum of gapless fluctuation modes $\omega_q^- \approx v_+q$, since the main contribution to the integral is in the limit of low $q$. The integration yields
\begin{align}
\Gamma^-(\bar{\omega}) = 2 \frac{\delta^{-1}v_-^2}{v_+^3}\frac{T}{|\bar{\omega}|} + O(\bar{\omega}^0).
\end{align}
The resulting effective action for the collective coordinate is now given by
\begin{align}
S_{eff}[\xi] = T\sum\limits_{\bar{\omega}} \left[\frac{M}{2}\bar{\omega}^2 + 2 \frac{\delta^{-1}v_-^2}{v_+^3} T |\omega|  \right]\xi_{-\bar{\omega}}\xi_{\bar{\omega}}.
\end{align}

\section{Green function of bosonic fields \label{app:path-integrations}}
Here we express the retarded Green function $\mathcal{G}^R$ of the bosonic field $\phi_-$ via the Green functions $\mathcal{D}^{R}$, $\mathcal{D}^{K}$  of the collective coordinate $\xi$. We use the path integral formulation of the Keldysh technique described e.g. in~Ref.~\onlinecite{KamenevLevchenkoAdvPhys2010}. It is convenient to represent the retarded Green function as
\begin{align}
\mathcal{G}^R = \mathcal{G}^{++} - \mathcal{G}^{+-}.
\label{eqn:GR-relation}
\end{align}

We introduce the path integral
\begin{align}
I_{\alpha \beta}(q,q') = \int \mathcal{D}\xi\; e^{-iq \xi(t_\alpha) -iq'\xi(t'_\beta)} e^{iS[\xi]},
\end{align}
where the Keldysh indices $\alpha$, $\beta$ denote whether the time is taken on the upper or on the lower branch of the Keldysh-Schwinger contour. 

The Green functions $\mathcal{G}_{\pm\pm}$ given by Eq.~(\ref{eqn:Green-path-integrals}) can therefore be represented as
\begin{align}
\mathcal{G}_{\alpha\beta}(x,t,x,t') = \int \frac{dqdq'}{(2\pi)^2}\; \phi_-(q)\phi_-(q') e^{i(q+q')x} I(q,q').
\end{align} 

The action $S[\xi]$ in the limit $L\to\infty$ is translational invariant. Hence, the path integral $I(q,q')$ must depend only on $q-q'$. We will show this explicitly. 

In order to perform the path integration we use the time-discretization $t_1\to-\infty$, $t_2$, \dots, $t_N=t_{N+1}\to+\infty$, \dots, $t_{2N} = -\infty$. The discrete version of the path integral $I(q,q')$ becomes then
\begin{align}
I(q,q') = \int \frac{d\xi_1 dp_1\dots d\xi_{2N}dp_{2N}}{(2\pi)^{2N}}e^{-iq\xi_n -iq'\xi_{\xi_{n'}}}e^{iS[\xi,p]},
\end{align}
where the action is given by
\begin{multline}
S[\xi,p] = \sum\limits_{k=1}^{2N-1} i(p_k-p_{k+1})\xi_k\\ -\sum\limits_{k=1}^{2N}iH(p_k)(t_{k+1}-t_k)\\ + i\xi_{2N}(p_{2N}-p_1)-\beta\frac{p_1^2}{2M}.
\end{multline}
The last two terms arise from the equilibrium density matrix $\rho_0$ at $t\to-\infty$,
\begin{align}
\left\langle p_1 \middle| \rho_0 \middle| p_{2N} \right\rangle = \int d\xi_{2N}\;e^{  i\xi_{2N}(p_{2N}-p_1) -\beta \frac{p_1^2}{2M} }. 
\end{align}	
We shift $\xi_k \to \xi_1 + \tilde{\xi}_k$, and then integrate out $\xi_1$, $p_1$, $\xi_{2N}$, and $p_{2N}$ to get
\begin{multline}
I(q,q') = 2\sqrt{\frac{MT}{\pi}} 2\pi\delta(q+q')\\\times \int \frac{\prod\limits_{k=2}^{2N-1}d\tilde{\xi}_k dp_k}{(2\pi)^{2N-2}} e^{-iq(\tilde{\xi}_n)-iq'(\tilde{\xi}_{n'})}e^{i\tilde{S}},
\end{multline}
with the new action
\begin{multline}
S[\tilde{\xi},p] = ip_2\tilde{\xi}_2 + \sum\limits_{k=2}^{2N-2} i\tilde{\xi}_k (p_{k+1}-p_k)\\ -\sum\limits_{k=2}^{2N-1}iH(p_k)(t_{k+1}-t_k)\\  -\beta\frac{M\xi_{2N-1}^2}{2(\delta t)^2}.
\end{multline}
The continuum version of the path integral then reads
\begin{align}
I(q,q') = \sqrt{\frac{MT}{\pi}}4\pi \delta(q+q')\int \mathcal{D}\tilde{\xi} e^{-iq\tilde{\xi}(t_\alpha)-iq'\tilde{\xi}(t'_\beta)}
e^{iS[\tilde{\xi}]}.
\end{align}
Now the integration over $\tilde{\xi}$ can be performed straightforwardly. We obtain
\begin{align}
I_{\alpha \beta}(q,q',t,t') = \sqrt{\frac{MT}{\pi}} 4\pi \delta(q+q')e^{-iq^2\left[\mathcal{D}_{\alpha\beta}(0)-\mathcal{D}_{\alpha\beta}(t-t')\right]},
\end{align}
and the Green function $\mathcal{G}_{\alpha \beta}$ is given by
\begin{align}
i\mathcal{G}_{\alpha\beta}(t) = 2\sqrt{\frac{MT}{\pi}}\int\frac{dq}{2\pi} |\phi_-(q)|^2 e^{-iq^2\left[\mathcal{D}_{\alpha\beta}(0)-\mathcal{D}_{\alpha\beta}(t)\right]}.
\end{align}
The Green functions $\mathcal{D}_{\alpha \beta}$ are related to retarded, advanced, and Keldysh functions by
\begin{align}
&\mathcal{D}_{++} = \frac{\mathcal{D}^R+\mathcal{D}^A+\mathcal{D}^K}{2}\label{eqn:Dpp-relation},\\
&\mathcal{D}_{+-} = \frac{\mathcal{D}^A-\mathcal{D}^R+\mathcal{D}^K}{2} \label{eqn:Dpm-relation}.
\end{align}

Using Eqs.~(\ref{eqn:GR-relation}),~(\ref{eqn:Dpp-relation})--(\ref{eqn:Dpm-relation}) we obtain
\begin{multline}
i\mathcal{G}^R(t)=\Theta(t)2\sqrt{\frac{MT}{\pi}}\int \frac{dq}{2\pi} 2i|\varphi(q)|^2e^{-iq^2\left[\mathcal{D}^K(0)-\mathcal{D}^K(t) \right]}\\ \times\sin\left(q^2\left[\mathcal{D}^R(0) - \mathcal{D}^R(t)\right] \right).
\end{multline}
Finally, integration over $q$ yields Eq.~(\ref{eqn:retarded}) in the main text.

\section{Temperature dependence of the order parameter \label{app:order}}

The value of the order parameter $m$ at given temperature $T$ is given by a self-consistent equation, see Eq.~11 of~Ref.~\onlinecite{MengEPLJB2014},
\begin{align}
m = B_I\left(\dfrac{\epsilon I}{T}\right),
\label{eqn:order}
\end{align}
with $B_I$ denoting the Brillouin function~\cite{Kittel1986}, and $\epsilon(m) = \epsilon_M+\epsilon_P+\epsilon_K$ consisting of magnon energy, Peierls-like energy gain and Knight-shift energy, respectively,
\begin{align}
&\epsilon_M = m \frac{I}{N_\perp} \mathcal{C}(g) \frac{A^2 a}{\hbar v'}\left(\frac{l'_\xi}{a}\right)^{2-2g},\\
&\epsilon_P = \frac{1}{\pi} m \frac{I}{N_\perp} \frac{A^2 a}{\hbar v'}\left(\frac{l'_\xi}{a}\right)^{2-2g} \ln\left(\frac{2\varepsilon_F}{m A I}\right),\\
&\epsilon_K = \frac{1}{2\pi} m \frac{1}{N_\perp} \frac{A^2 a}{\hbar v'}\left(\frac{l'_\xi}{a}\right)^{2-2g} \ln\left(\frac{2\varepsilon_F}{m A I}\right),
\end{align}
where $g = K_\rho\sqrt{\dfrac{2}{1+K_\rho^2}}$, with renormalized velocity $v' = \dfrac{v_F}{K_\rho} \sqrt{\dfrac{1+K_\rho^2}{2}}$, renormalized correlation length $l'_\xi = \min\{L, \hbar v'/T, \hbar v'/ \Delta \}$, and the dimensionless factor $\mathcal{C}$ is given by
\begin{align}
\mathcal{C}(g) = \frac{\sin(\pi g)}{2}\left( 2\pi \right)^{2g-4} \Gamma^2(1-g)\left|\frac{\Gamma(g/2)}{\Gamma(1-g/2)}\right|.
\end{align}
Here, $\Gamma(g)$ denotes the gamma function.

At temperatures near the critical $T_c$, where $m \ll 1$, the Brillouin function $B_I$ can be linearized, $B_I(x) \approx \dfrac{1+I}{3I} x$, and the self-consistent condition~(\ref{eqn:order}) can be rewritten as
\begin{align}
1 = \left(\frac{T_c}{T}\right)^\nu\left[ 1 + \zeta \ln \frac{2\epsilon_F}{m A I} \right],
\label{eqn:log-order}
\end{align}
where the factor $\zeta$ is given by
\begin{align}
\zeta = \frac{2I+1}{2\pi I \mathcal{C}(g)},
\label{eqn:zeta}
\end{align}
and the exponent $\nu$ depends on whether the correlation length is determined by the temperature or by the finite length of the wire
\begin{align}
\nu=\left\{
\begin{aligned}
&1,&\varepsilon'_L=\frac{\hbar v'}{L} > T_c,\\
&3-2g,&\varepsilon'_L =\frac{\hbar v'}{L} < T_c
\end{aligned}
\right.
\label{eqn:lambda}.
\end{align}
Solving~Eq.~(\ref{eqn:log-order}) we obtain the dependence of the order parameter above  $T_c$,
\begin{align}
m \propto \exp\left\{ - \frac{1}{\zeta} \left(\frac{T}{T_c} \right)^\nu \right\}.
\end{align}


\end{document}